\documentclass[12pt,preprint]{aastex}
\usepackage{rotating}
\usepackage{graphicx}
\usepackage{color}

\slugcomment{The Astronomical Journal: in press}

\shorttitle{313P/Gibbs}
\shortauthors{Jewitt et al.}

\begin{document}

\title{Nucleus and Mass Loss from Active  Asteroid 313P/Gibbs}
\author{David Jewitt$^{1,2}$,  Jing Li$^{1}$, Jessica Agarwal$^3$, Harold Weaver$^4$,  Max Mutchler$^5$, and Stephen Larson$^6$
}
\affil{$^1$Department of Earth, Planetary and Space Sciences,
UCLA, 595 Charles Young Drive East, 
Los Angeles, CA 90095-1567\\
$^2$Department of Physics and Astronomy,
University of California at Los Angeles, \\
430 Portola Plaza, Box 951547,
Los Angeles, CA 90095-1547\\
$^3$Max Planck Institute for Solar System Research, Max-Planck-Str. 2, 37191 Katlenburg-Lindau, Germany\\
$^4$The Johns Hopkins University Applied Physics Laboratory, 11100 Johns Hopkins Road, Laurel, Maryland 20723  \\
$^5$Space Telescope Science Institute, 3700 San Martin Drive, Baltimore, MD 21218 \\
$^6$Lunar and Planetary Laboratory, University of Arizona, 1629 E. University Blvd.
Tucson AZ 85721-0092 \\
}

\email{jewitt@ucla.edu}

\begin{abstract}
We present Hubble Space Telescope observations of active asteroid 313P/Gibbs (formerly P/2014 S4) taken over the five month interval from 2014 October to 2015 March.  This object has been recurrently active near perihelion (at 2.4 AU) in two different orbits, a property that is naturally explained by the sublimation of near surface ice but which is difficult to reconcile with other activity mechanisms.   We find that the mass loss peaks near 1 kg s$^{-1}$ in October and then declines over the subsequent months by about a factor of five, at nearly constant heliocentric distance.  This decrease is too large to be caused by the change in heliocentric distance during the period of observation.  However, it is consistent with  sublimation from an ice patch shadowed by local topography, for example in a pit like those observed on the nuclei of short-period comet 67P/Churyumov-Gerasimenko.  While no unique interpretation is possible, a simple self shadowing model shows that  sublimation from a pit with depth to diameter ratio near 1/2 matches the observed rate of decline of the activity, while deeper and shallower pits do not.  We estimate the nucleus radius to be 700$\pm$100 m (geometric albedo 0.05 assumed).   Measurements of the spatial distribution of the dust were obtained from different viewing geometries.  They show that dust was ejected continuously not impulsively, that the effective particle size is large, $\sim$50 $\mu m$, and that the ejection speed is $\sim$2.5 m s$^{-1}$.  The total dust mass ejected is $\sim$10$^7$ kg, corresponding to $\sim$10$^{-5}$ of the nucleus mass.  The observations are consistent with partially shadowed sublimation from $\sim$10$^4$ m$^2$ of ice, corresponding to $\sim$0.2\% of the nucleus surface.  For ice to survive in 313P for billion-year timescales requires that the duty cycle for sublimation be $\lesssim$10$^{-3}$.
\end{abstract}


\section{INTRODUCTION}
313P/Gibbs (formerly P/2014 S4), hereafter called ``313P'', was discovered on UT 2014 September 24 as a product of the on-going Catalina Sky Survey (Gibbs 2014). Although 313P is cometary in appearance, its orbit lies within the asteroid belt, having a semimajor axis 3.156 AU, eccentricity 0.242 and inclination 11.0\degr.  Perihelion and aphelion occur at 2.392 AU and 3.920 AU, respectively. The Tisserand parameter with respect to Jupiter, $T_J$ = 3.13, is distinct from those of Kuiper belt and Oort cloud comets, which have 2 $\le T_J \le$ 3 and $T_J <$ 2, respectively (Kresak 1982).   The simultaneous comet-like appearance and asteroid-like orbit establish 313P as a member of the active asteroids population.  

The active asteroids are driven by a variety of mechanisms, from impact, to sublimation, to thermal fracture and to rotational instabilities, all previously thought to lie beyond the realm of observation (Jewitt 2012).  In a majority of objects, the observational constraints are sufficiently limited that no single activity mechanism can be uniquely identified.  However, observations show that 313P was also active near perihelion in archival observations taken two orbits earlier in 2003,  but not in 2004 (Hsieh et al.~2015, Hui and Jewitt 2015, Jewitt et al.~2015).  This establishes 313P as one of only three main-belt bodies, along with 133P/Elst-Pizarro (Hsieh et al.~2004) and 238P/Read (Hsieh et al.~2011), to display activity recurrent on different orbits. Repetition in different orbits immediately rules out most of the possible activity mechanisms and leaves sublimation as the most likely remaining candidate (c.f.~Hsieh and Jewitt 2006).  In support of this conclusion, the distribution of surface brightness in data from the active epochs in both 2003 and 2014 is best matched by protracted (as opposed to impulsive) dust emission, as expected from comet-like sublimation.  Even though a search for direct evidence of outgassing via the CN resonance fluorescence lines proved negative (setting an upper limit to the gas production rate $<$1.8 kg s$^{-1}$; Jewitt et al.~2015), the measured properties strongly suggest an origin by sublimation. 

We initiated a program of observations designed to characterize 313P and to determine the origin of its activity; here we present our initial results.

\section{OBSERVATIONS} 
Initial observations using the 2.4 meter diameter Hubble Space Telescope (HST) were obtained from two orbits allocated under program GO 13864. Based on these observations, an additional six orbits were awarded as Director's Discretionary Time under program GO 14040.  All measurements were taken using the WFC3 camera and the broadband F350LP filter (full width at half-maximum (FWHM) = 4758\AA), having an effective central wavelength 6230\AA~when used to observe a solar-type source.  The WFC3 camera consists of two charge-coupled devices each 2051$\times$4096 pixels, with square pixels 0.04\arcsec~on a side, giving a field of view 162\arcsec$\times$162\arcsec.   The image scale ranged from 42 km pixel$^{-1}$ on UT 2014 October 14 to 90 km pixel$^{-1}$ on UT 2015 March 05, increasing in proportion to the geocentric distance.     

We sampled the object at roughly monthly intervals in order to follow its morphological and photometric development.  Observations on UT 2015 January 07 were targeted to coincide with the passage of the Earth through the projected orbital plane of 313P.  From this special viewing geometry, the perpendicular extent of the dust can be interpreted free from the ambiguous effects of projection.   

A preliminary discussion of the data from GO 13864 is given in  Jewitt et al.~(2015).  Here, we present a full analysis of 313P from the combined HST datasets.   A brief log of the observations is  presented in Table \ref{geometry}.

 \section{RESULTS} 
 
 \subsection{Morphology}
 \label{morpho}
\begin{figure}
\epsscale{0.95}
\begin{center}
\plotone{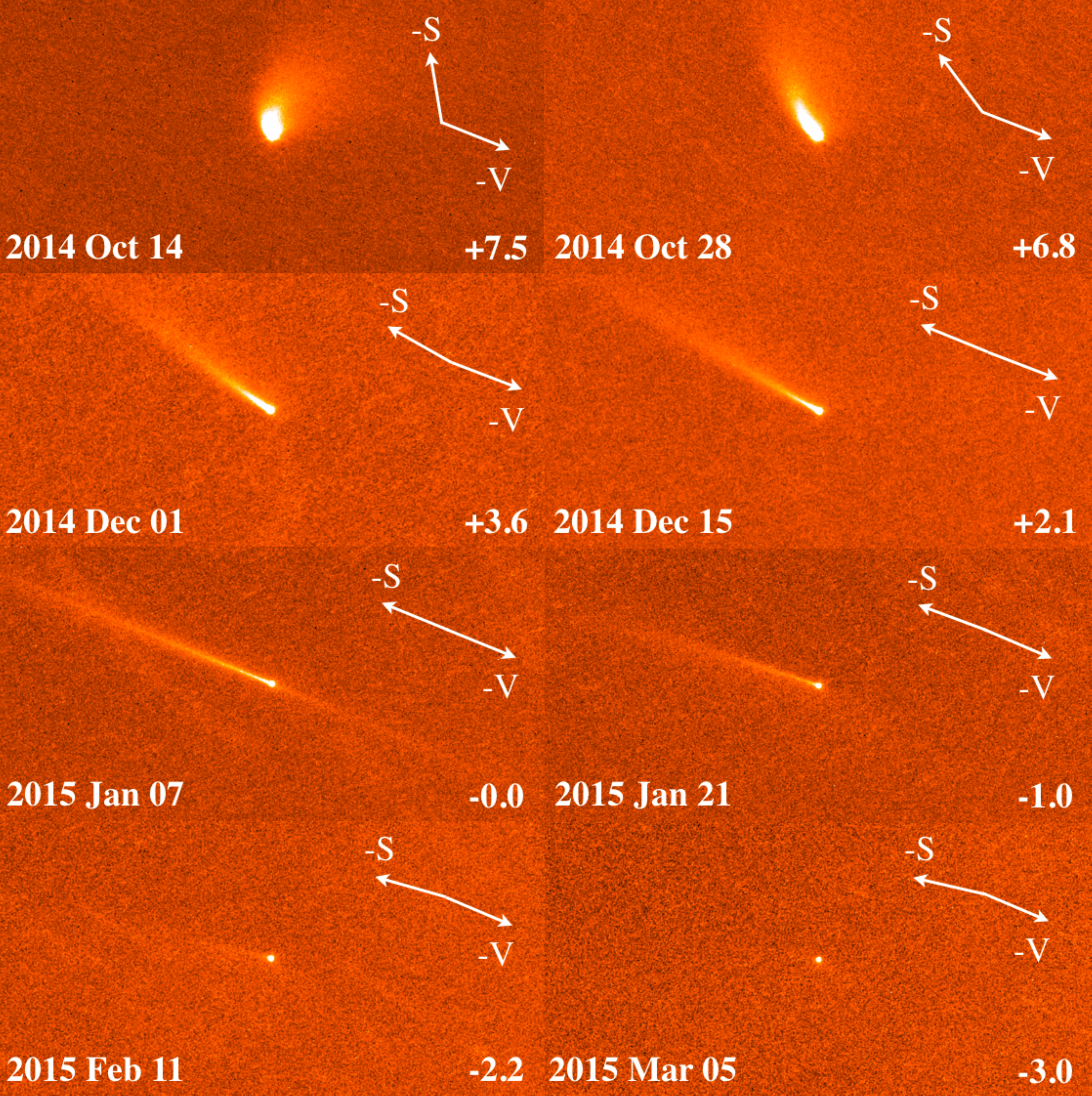}
\caption{Composite of HST images of 313P/Gibbs in which each panel has North to the top, East to the left, and is 30\arcsec~wide.  The dates of the images are indicated.  Arrows show the directions of the projected antisolar vector (-S) and the negative projected heliocentric velocity vector (-V).  Numbers in the lower right corner of each panel show the angle of the Earth above the orbital plane of the object, in degrees.   \label{composite}
} 
\end{center} 
\end{figure}

Figure (\ref{composite}) shows a composite of the HST images at fixed angular scale (each panel is 30\arcsec~in width), with direction vectors showing the anti-solar and negative heliocentric velocity directions.  To construct each panel, the images from each orbit of HST were shifted and combined to eliminate cosmic rays and other defects that are present in the individual images.  In most cases, the cosmic ray removal was successful but residual large scale features caused by the parallactic smearing of field stars and galaxies sometimes remain.  For example, the composite from UT 2015 January 07 shows a diffuse background arc that crosses the tail to the east of the nucleus and which cannot be removed by digital processing.  Several trailed galaxies are also evident in the composite from UT 2015 February 11.  Such incompletely removed background sources limit the accuracy of surface photometry but, in most cases, do not detract from the conclusions to be reported here. 

The morphology of 313P changes systematically with time from showing a nucleus plus bright, fan-shaped tail in the early observations from 2014 October (discussed in Paper 1) to a thin, linear tail as the Earth passes through the orbital plane in January, towards an increasingly point-like appearance until the last observation on 2015 March 05 (Figure \ref{composite}).  The position angle of the tail swings counter-clockwise, following the progression of the projected anti-solar vector from panel to panel in the figure.  There is no evidence for a resolved coma about the nucleus.  While the March 05 image at first appears stellar, very faint extended emission is present in the antisolar direction, showing that the nucleus was not completely bare. Nevertheless, it is clear from  Figure (\ref{composite}) that much less dust is present in the later images of the sequence than in the earlier ones.  

\begin{figure}
\begin{center}
\includegraphics[width=1.10\textwidth, angle =270 ]{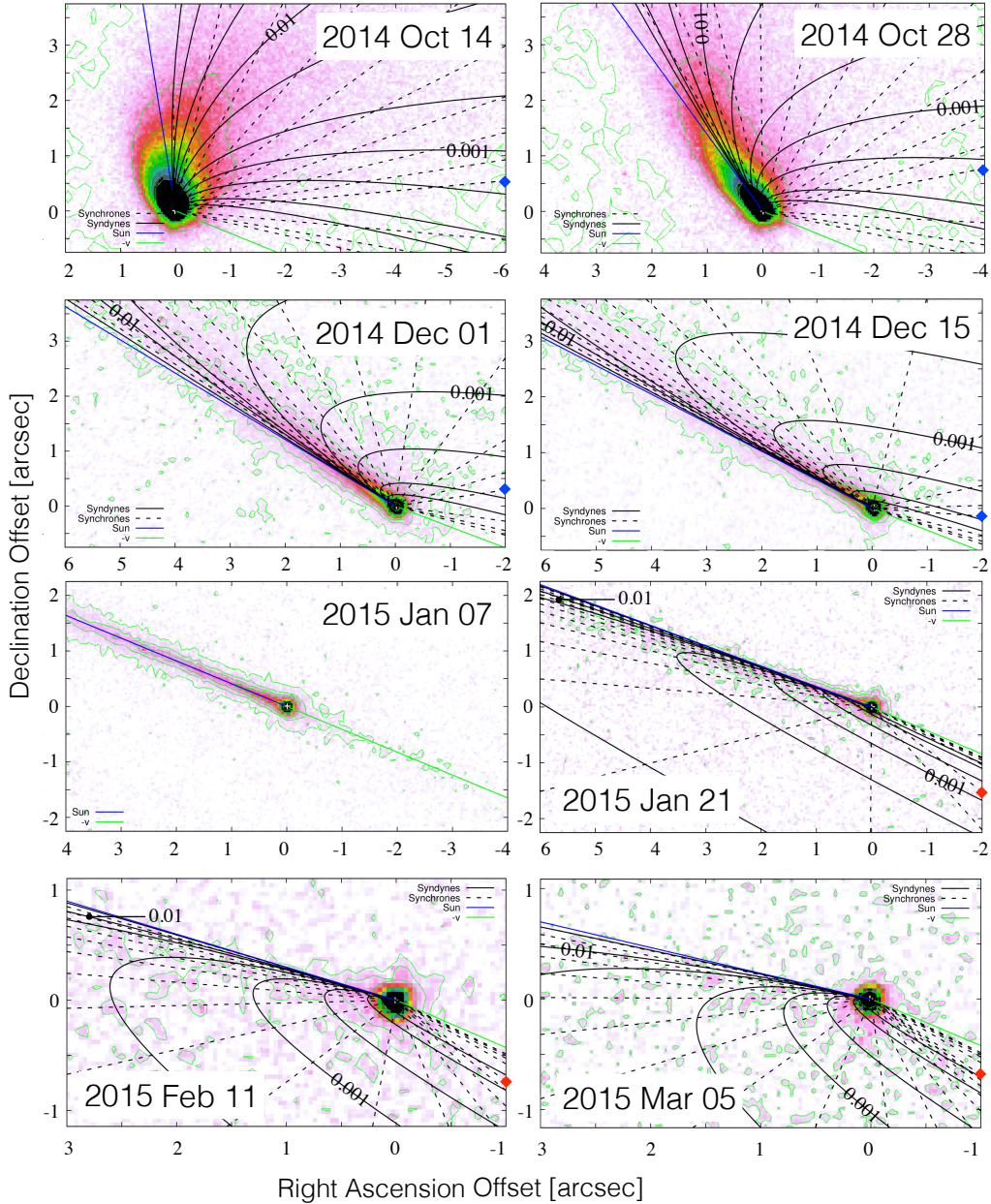}

\caption{Syndynes (solid black lines)  for $\beta$ = 0.1, 0.05, 0.02, 0.01, 0.005, 0.002, 0.001, 0.0005, 0.0002 and 0.0001 are shown. However, only the $\beta$ = 0.01 and 0.001 syndynes are  labeled for clarity.   Labels for synchrones (dashed straight lines) are also suppressed for clarity but can be determined as follows.   The blue diamond corresponds to ejection on 2014-July-9, with sychrones plotted every 10 days afterwards (anti-clockwise) and
50 days before. The red diamond is for ejection on 2014-Aug-28 and plotting intervals are the same.   \label{comp_models}
} 
\end{center} 
\end{figure}

We computed syndyne/synchrone (Finson and Probstein 1968) models for each of the dates of observation in Table (\ref{geometry}).  These models are shown in Figure (\ref{comp_models}), where solid black lines are used to indicate syndynes (the loci of positions of particles of one size released from the nucleus over a range of times) and dashed black lines to indicate synchrones (loci of particles having a wide range of sizes released at a given time).  We note that for 2015 January 07, all synchrones and syndynes are collapsed onto the projected orbit, with  synchrones  later than 2014 September 03  to the East of the
nucleus, and all earlier synchrones to the west.  The dust particles are assumed to be released from the nucleus at zero initial speed and then to be accelerated by solar radiation pressure with value $\beta g_{\odot}$, where $\beta$ is dimensionless and $g_{\odot}$ is the local gravitational acceleration to the Sun.   For particles whose circumference is larger than the wavelength of light, $2\pi a > \lambda$, it may be shown that $\beta \sim a_{\mu m}^{-1}$, where $a_{\mu m}$ is the radius expressed in microns (Bohren and Huffman 1983). 

Figure (\ref{comp_models}) shows that the curvature of the dust tail and the sky-plane rotation of the tail with respect to time are both naturally reproduced by the syndyne trajectories of particles having characteristic $\beta \sim$ 0.02, or size $a \sim$ 50 $\mu m$.  The same conclusion was reached by Jewitt et al.~(2015), Hsieh et al.~(2015), Hui and Jewitt (2015) and Pozuelos et al.~(2015).  The failure of synchrone trajectories (straight lines in Figure \ref{comp_models}) to match the tail isophotes is evidence that the dust emission from this body is unlikely to be caused by impact or any other impulsive mechanism.  

\subsection{Photometry}

We measured photometry of 313P in two ways.  The brightness within a set of projected circular apertures of fixed angular radii is listed in Table (\ref{photometry}).    Photometry of a distributed source using fixed angular radius apertures is affected by the geocentric distance, $\Delta$, because the volume of the coma sampled by each aperture increases with the cube of increasing $\Delta$.  Therefore, we also determined the brightness within a set of apertures scaled to have fixed linear radius at the instantaneous distance of the object.  These measurements are summarized in Table (\ref{absolute}).  The sky level for all measurements was determined from the median count within a concentric ``sky'' annulus extending from 6\arcsec~to 12\arcsec.   As discussed in Section (\ref{sb_profiles}) the coma surface brightness falls quickly with increasing distance from the nucleus, and background subtraction from  the sky annulus was found to be not critical to the photometry.  

\begin{figure}
\epsscale{0.85}
\begin{center}
\plotone{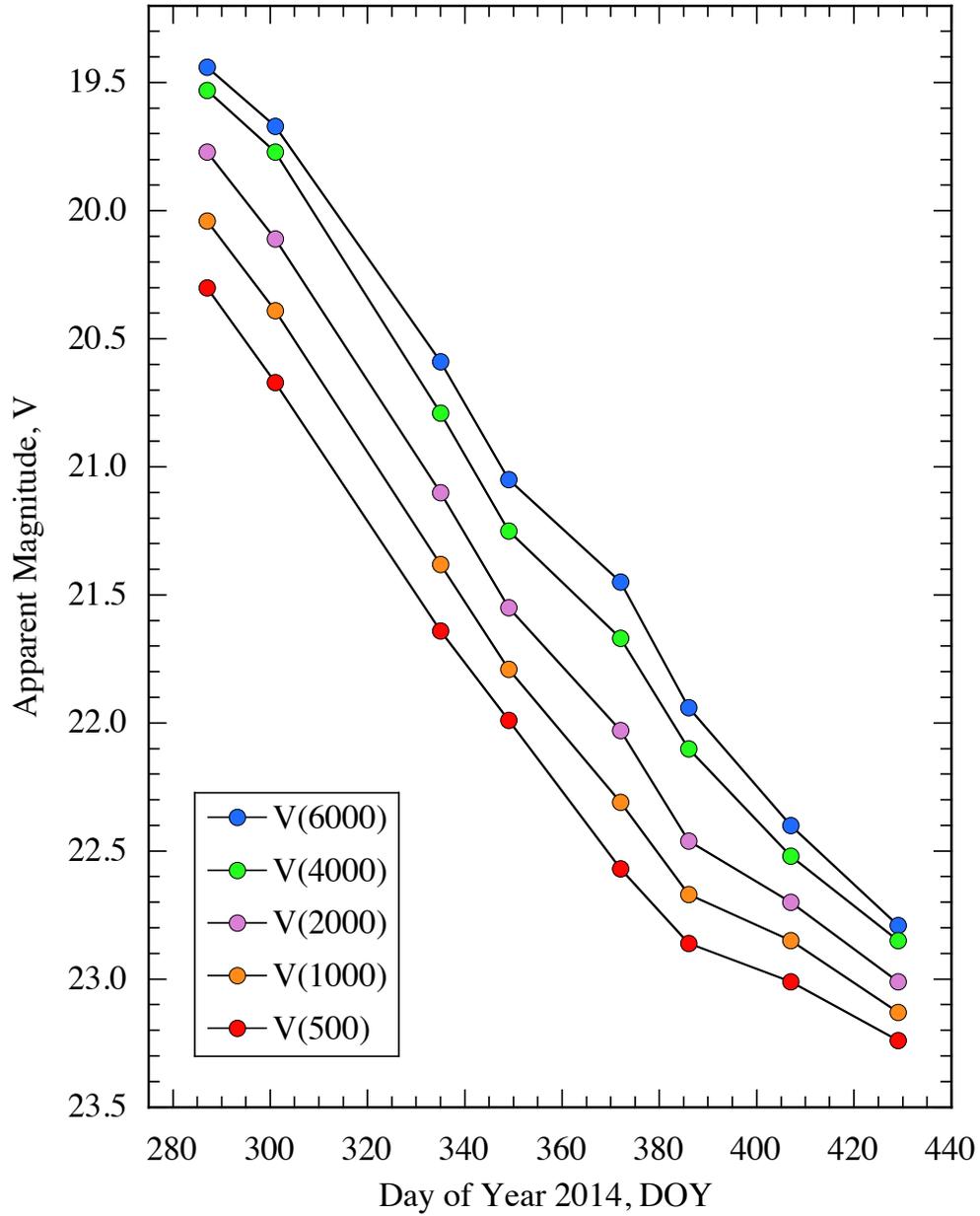}
\caption{Apparent magnitudes determined within circular apertures of fixed outer radii, from 500 km to 6000 km, as marked.  The statistical measurement error bars ($\pm$0.01 magnitude) are too small to be seen at the scale of the plot.  \label{V_vs_DOY}
} 
\end{center} 
\end{figure}

The brightness of 313P shows a steady decline with time.  In Figure (\ref{V_vs_DOY}) we plot the brightness within a set of concentric photometry apertures having fixed linear radii, from 500 km to 6000 km, projected to the distance of 313P.   The Figure shows that the apparent brightness of 313P faded  by $\sim$3 magnitudes between the first measurement on October 14 and the last on March 05.  Some of the fading is caused by the increasing distance and the larger phase angle of the later observations (Table \ref{geometry}).  

\begin{figure}
\epsscale{0.85}
\begin{center}
\plotone{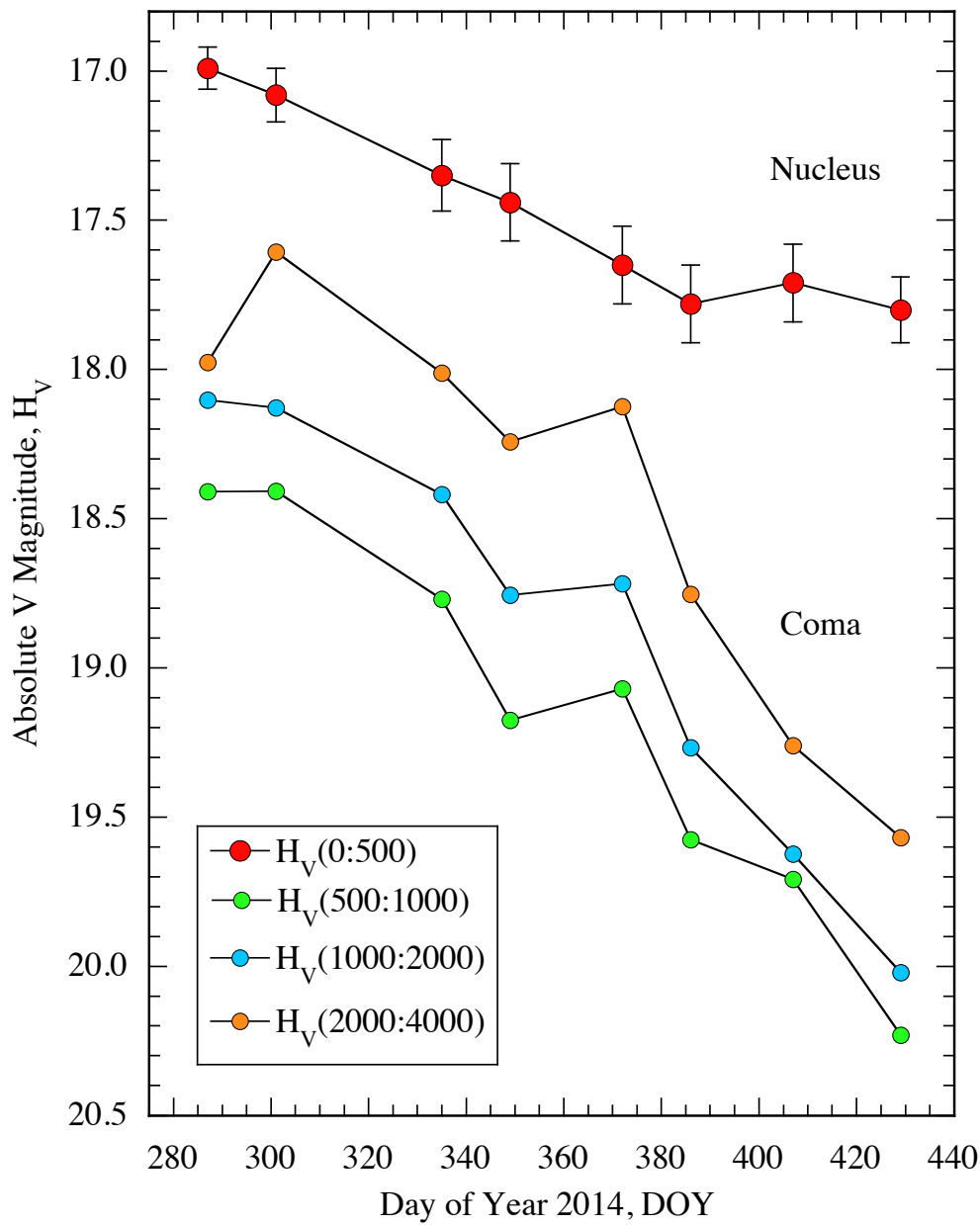}
\caption{Absolute magnitudes determined within circular annuli scaled to have fixed linear inner and outer radii, as indicated, in kilometers.  The error bars on the smallest annulus, $H_V(0$:$500)$, show systematic uncertainties resulting from the unmeasured phase function of 313P.  Similar uncertainties exist for the larger annuli  but are not shown to avoid plot confusion.  \label{Hv_vs_DOY}
} 
\end{center} 
\end{figure}

In Figure \ref{Hv_vs_DOY} we have corrected the apparent magnitudes to ``absolute'' magnitudes using 

\begin{equation}
H_V = V - 5\log(R\Delta) + 2.5\log_{10}(\Phi(\alpha)).
\label{absolute}
\end{equation}

\noindent Here, $\Phi(\alpha)$ is the phase function at phase angle $\alpha$, equal to the ratio of the scattered light at $\alpha$ to that at $\alpha$ = 0\degr.   We assumed the phase function formalism of Bowell et al.~(1989) with parameter $g$ = 0.15, as appropriate for a C-type object, as suggested by the color of 313P and by its orbital location in the outer asteroid belt.  The phase function of 313P is unmeasured, introducing an uncertainty into the value of $H_V$ that is larger than the ($\sim$0.01 magnitude) uncertainty of the photometry.   To estimate the possible size of this uncertainty, we set the error on $H_V$ to be equal to the difference between the phase function corrections for C-type and S-type objects.   Over the range of phase angles at which 313P was observed, this difference is typically $\sim$0.1 magnitudes.   Absolute magnitudes are given in Table (\ref{photometry}) with their statistical uncertainties.  

The absolute magnitudes are related to the effective scattering cross-section of the material within the photometry aperture, $C_e$ (km$^2$),  by

\begin{equation}
C_e = \frac{2.24\times10^{16} \pi}{p_{V}} ~10^{0.4[m_{\odot, V} - m_{V}(1,1,0)]}
\label{area}
\end{equation}

\noindent where $p_{V}$ is the geometric albedo of 313P and $m_{\odot, V}$ is the apparent magnitude of the Sun, both at the wavelength of the V filter.  We assume $m_{\odot, V}$ = -26.77.  The resulting scattering cross-sections are listed in Table (\ref{photometry}), computed assuming $p_{V}$ = 0.05 (c.f.~Fernandez et al.~2013).  Uncertainties on $C_e$ are systematic in nature, dominated by the phase function correction as well as by the assumption of the geometric albedo.

The absolute brightness of the coma annuli in Figure (\ref{Hv_vs_DOY}) decreases until the last observation on March 05 (DOY 429), but the central aperture flattens after UT 2015 January 21 (DOY 386) and, within the uncertainties of measurement, remains constant.  The central aperture is most strongly sensitive to the brightness of the nucleus.  The effective cross-section within the 500 km radius aperture is $C_e$ = 2.2$\pm$0.1 km$^2$ (Table \ref{absolute}).  Formally, this gives an upper limit to the nucleus cross-section because there remains a coma even in the latest observations (see Figure \ref{composite}).  The corresponding value of the effective circular radius of the nucleus is $r_n = (C_e/\pi)^{1/2}$ = 0.8 km.

\subsection{Radial Surface Brightness Profile}
\label{sb_profiles}
\begin{figure}
\epsscale{0.95}
\begin{center}
\plotone{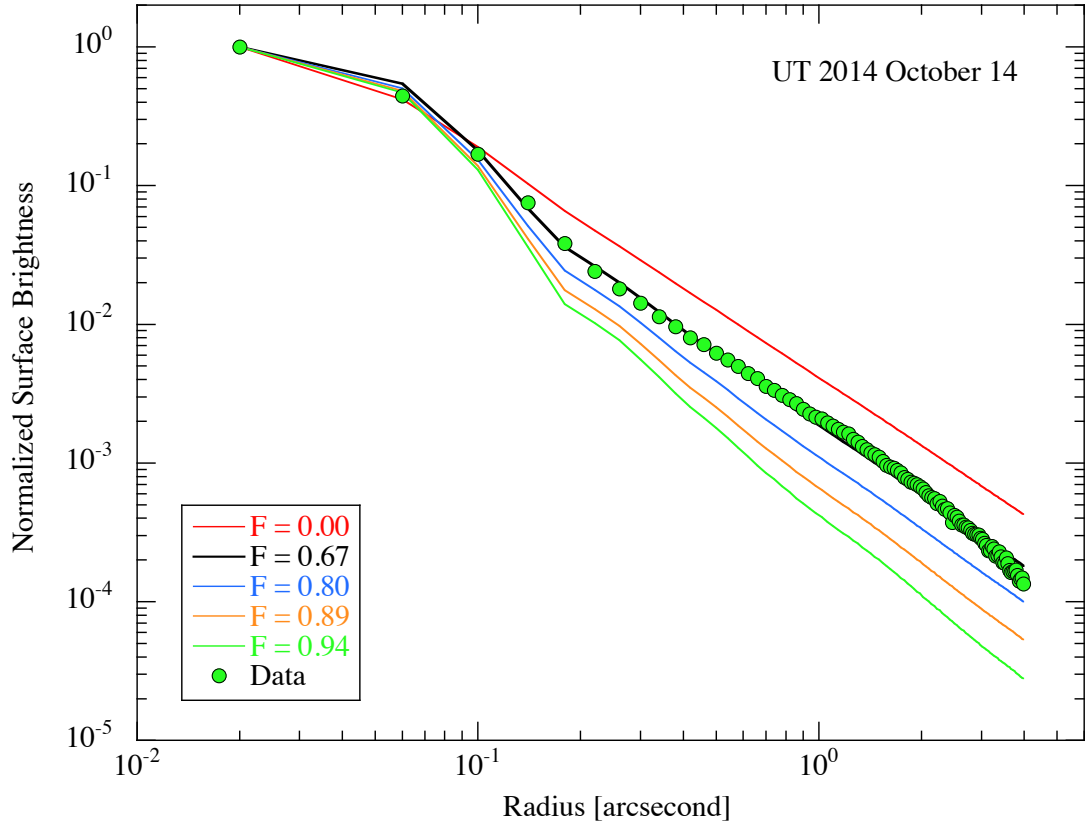}
\caption{Normalized surface brightness profiles of 313P measured on UT 2014 October 14 (green circles) compared with models. Quantity $F$ is the fraction of the signal within angular radius $\theta \le$ 0.2\arcsec~that is contributed by scattering from the nucleus (see Equation \ref{convolve}).  \label{sb_oct14}
} 
\end{center} 
\end{figure}

\begin{figure}
\epsscale{0.95}
\begin{center}
\plotone{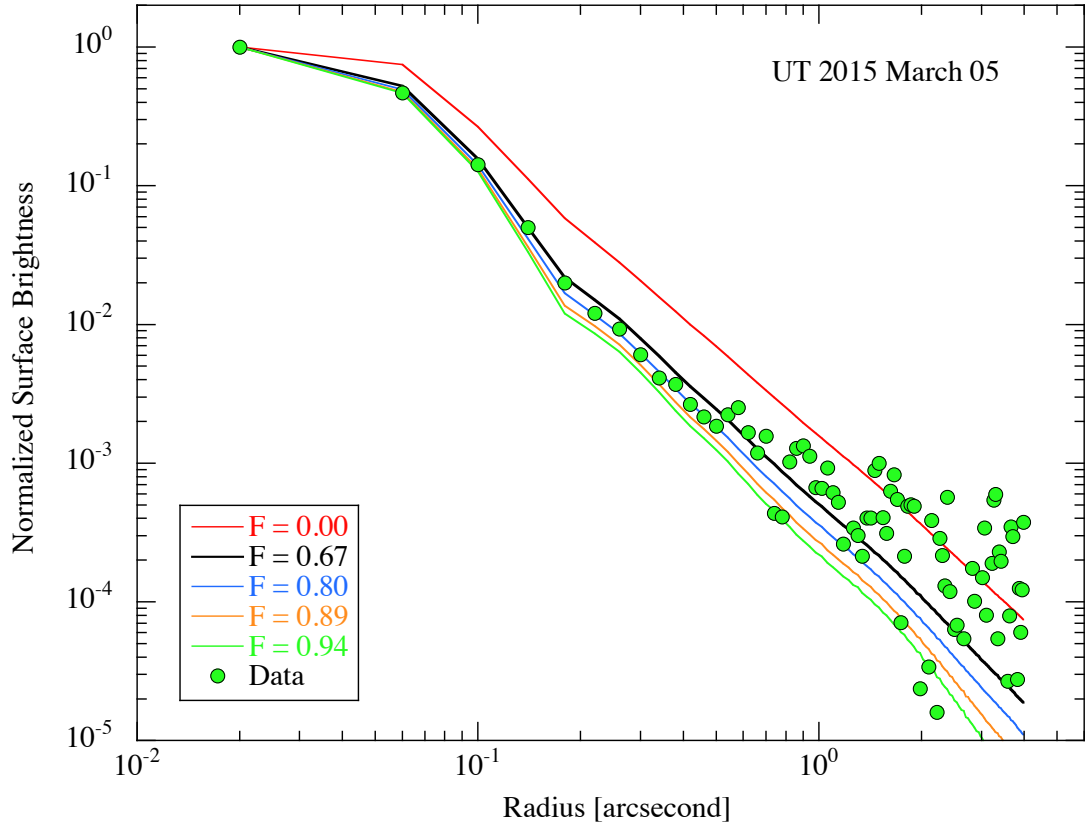}
\caption{Same as Figure (\ref{sb_oct14}) but for observations taken UT 2015 March 05. \label{sb_mar05}
} 
\end{center} 
\end{figure}

We measured the surface brightness, $\Sigma(\theta)$, as a function of the angular radius, $\theta$, for each epoch of observation within a set of concentric annuli centered in the nucleus.  For this purpose, we first shifted the individual images within each HST orbit to be aligned on the nucleus. Cosmic rays and background sources were then eliminated by computing the median data number within each pixel from each date.   Sky background subtraction was determined from the median signal measured in a concentric annulus with inner and outer radii $\theta$ = 6.0\arcsec~and 12\arcsec, respectively.  Differences between the profiles, when normalized to unity at the peak, are modest and correspondingly difficult to illustrate graphically.  We show the profiles on UT 2014 October 14 (Figure \ref{sb_oct14}) and 2015 March 05 (Figure \ref{sb_mar05}), these being the first and last dates of observation and illustrating the largest profile differences.  The noise on the profiles is indicated by the scatter between measurements in adjacent annuli, with the profile on March 05 being very noisy as a result of the greater distance and secular fading of the coma.  

%

The surface brightness profiles all display a central bump at angular radii $\theta \le$ 0.2\arcsec, caused by the combined effects of scattering from the nucleus and from the coma, all convolved with the point-spread function (PSF).   Our main interest in the radial surface brightness profiles is to estimate the effective cross-section (and size) of the nucleus.  In principle, the nucleus and dust contributions can be disentangled by deconvolution, using the known PSF.  In practice, however, deconvolution acts as a noise amplifier, and we instead elected to estimate the nucleus contribution to the surface brightness profile by convolving simple models of the intrinsic profile with the instrumental PSF.  To construct the models, we assumed that the intrinsic profile can be represented by a centrally located, unresolved nucleus, represented by a single pixel of strength $\delta_n(\theta)$, and a coma in which the surface brightness varies with angular radius $\theta$ as $\Sigma(\theta) = k \theta^{-p}$, where $k$ is a constant representing the strength of the coma.  We solved for index $p$ by fitting the observed surface brightness profiles in the radius range 0.2\arcsec $\le \theta \le$ 1.0\arcsec, this being large enough to avoid the central bump in the radial surface brightness.    Values of $p$ are listed together with their formal, least-squares fit uncertainties in Table (\ref{profile_fit}).  The actual uncertainties are larger than the listed formal errors because, even within the tiny 0.2\arcsec~to 1.0\arcsec~range, deviations from power-law behavior are apparent.  For example, the fit to data from 2015 February 11 was rendered particularly uncertain by structure inside $\theta$ = 1\arcsec~likely due to imperfectly removed cosmic rays and is omitted from further discussion.   

The model is computed from the convolution

\begin{equation}
\Sigma(\theta) = PSF(\theta) \ast (\delta_n(\theta) + k \theta^{-p}).  
\label{convolve}
\end{equation}

\noindent Here, $PSF(\theta)$ is the two-dimensional representation of the PSF computed using TINYTIM software (Krist et al.~2011), $\delta_n(\theta)$ is the Dirac delta function to represent the nucleus.   All convolution calculations were done at a resolution 0.01\arcsec~(pixel)$^{-1}$ and then rebinned to the WFC3 resolution of 0.04\arcsec~(pixel)$^{-1}$ to compare with the data.

Models were parametrized in terms of the quantity $F$, defined by

\begin{equation}
F = \frac{\delta_n(\theta)}{\delta_n(\theta) + 2\pi k \int_0^{0.2} \theta^{(1-p)}d\theta}.
\label{convolve}
\end{equation}

\noindent Quantity $F$ gives the ratio of the light scattered from the nucleus to the total  light (nucleus plus coma) scattered within the aperture of angular radius $\theta \le$ 0.2\arcsec.  $F$ = 0 corresponds to no nucleus contribution while $F = 1$ corresponds to a bare nucleus.  The denominator of Equation (\ref{convolve}) is proportional to the fixed-angle photometry summarized in Table (\ref{photometry}).


Results from the convolution models are summarized in Table (\ref{profile_fit}).  Even models in which the nucleus is assumed to be negligible ($F$ = 0) display a central surface brightness bump, owing to smearing of the coma brightness by convolution with the PSF.  However, all successful fits to the data require the presence of a nucleus (i.e.~$F >$ 0) in order to match the height of the central bump.  We obtained values of $F$ scattered in the range 0.6 $\le F \le $ 0.9, as listed in Table (\ref{profile_fit}).  The Table lists the nucleus cross-section computed from $C_n = F C_{0.2}$, where $C_{0.2}$ is the cross-section within the central 0.2\arcsec~taken from Table (\ref{photometry}). The final column of Table (\ref{profile_fit}) lists the effective nucleus radius in kilometers, computed from $r_n = (C_n/\pi)^{1/2}$.  The uncertainties on $F$ are estimated from the model fits to the data.  Uncertainties on $C_n$ and $r_n$ were calculated by propagation of errors.  In all cases, the uncertainties should be regarded as approximate.  As before, all listed cross-sections are predicated on the assumption of geometric albedo $p_V$ = 0.05.  

Table (\ref{profile_fit}) shows that the effective nucleus radius varies in the range $0.6 \le r_n \le 0.8$ km.  We do not know if this variability reflects uncertainties introduced by approximations in the method of coma removal, or by real variations associated with the rotation of an aspherical nucleus, or by a combination of the two.    However, we note that the fitted values of $C_n$ in column 5 of the Table are broadly consistent with the value $C_n$ = 2.0 km$^2$ obtained from the 500 km aperture photometry in UT 2015 March (Table \ref{absolute}).  The sum of the evidence suggests that 0.7 km is a reasonable measure of the nucleus radius.  With density $\rho$ = 10$^3$ kg m$^{-3}$, the escape velocity from a non-rotating, spherical 0.7 km radius nucleus is $V_e$ = 0.5 m s$^{-1}$.

\subsection{Perpendicular Profile}
\begin{figure}
\epsscale{0.95}
\begin{center}
\plotone{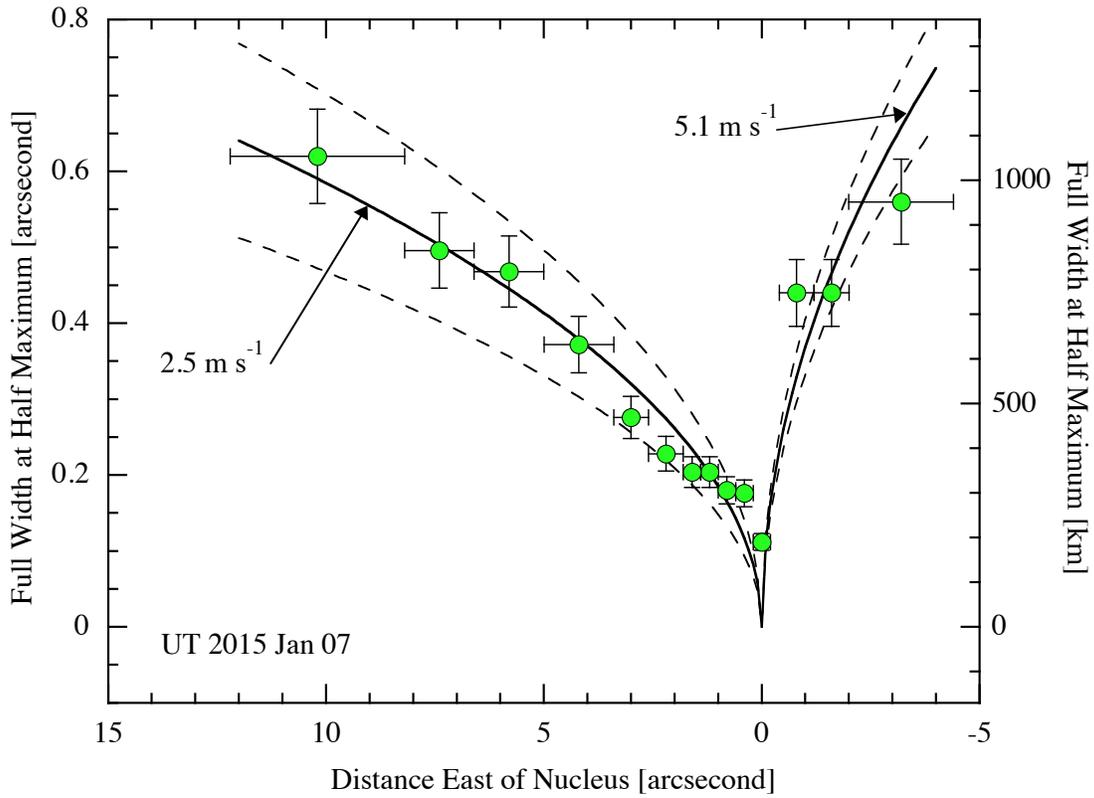}
\caption{Full width at half maximum of the dust emitted from 313P observed on UT 2015 January 07 from a vantage point in the orbital plane of the object.  Lines to the east mark ejection velocities 2.5$\pm$0.5 m s$^{-1}$ and to the west 5.1$\pm$0.5 m s$^{-1}$.  \label{FWHM}
} 
\end{center} 
\end{figure}

The UT 2015 January 07 composite image was used to determine the surface brightness profile of the dust in the direction perpendicular to the orbit.  With the Earth in the orbital plane on this date, the perpendicular extent accurately represents the vertical thickness of the dust distribution.  The resulting FWHM measurements are presented in Figure (\ref{FWHM}).  There, vertical error bars denote $\pm$10\% photometric uncertainties in the width determination, while horizontal bars mark the width of the dust tail segment used to obtain the measurement, the latter increasing with distance from the nucleus owing to the low surface brightness of the dust.  The dust projected to the west of the nucleus is particularly faint and therefore difficult to measure. As a result we show only the three FWHM measurements to the west in which we have confidence.  Likewise, although the dust is visible in Figure (\ref{composite}) more than 10\arcsec~to the east of the nucleus, we were unable to obtain convincing FWHM measurements beyond this distance owing to the low dust surface brightness and interference by a smeared galaxy (faintly visible in the figure as an oblique arc on the left hand side of the UT 2015 January 07 panel in Figure \ref{composite}).

The dust trail in Figure (\ref{FWHM}) is narrow, indicating that the dust particles are ejected at low velocity.  For dust particles ejected by gas drag the terminal velocity is proportional to the inverse square-root of the particle size, $V \propto a^{-1/2}$. In this case, it can be shown  that the width of the resulting trail, $w_T$, is related to the distance from the nucleus, $\ell_T$, by

\begin{equation}
V_{\perp} = \left(\frac{g_{\odot} w_T^2}{8 \ell_T}\right)^{1/2}
\label{thickness}
\end{equation}

\noindent  where $V_{\perp}$ is the component of the ejection velocity measured in the direction perpendicular to the orbit and $g_{\odot}$ is the local solar gravitational acceleration (Jewitt et al.~2014b).  For simplicity, we assume that $\ell_T$ is proportional to $\theta$,  the angular distance from the nucleus measured in the plane of the sky.   Two curves in Figure (\ref{FWHM}) show Equation (\ref{thickness}) with $V_{\perp}$ = 2.5 m s$^{-1}$ (on the east side of the nucleus) and $V_{\perp}$ = 5.1 m s$^{-1}$ (on the west side).  The assumption that   $\ell_T \propto \theta$ neglects projection effects, and is less accurate for the much older particles to the west of the nucleus, as may be seen in the curved syndynes of Figure \ref{comp_models}).  As a result, we take $V_{\perp}$ = 2.5 m s$^{-1}$  as the better estimate of the dust ejection velocity. For a different size-dependence of $V_{\perp}$, the width $w_T$ becomes a function of the particle size.  The measured trail width in this case can be taken as applicable at the optically dominant grain size which, from morphological analysis (section \ref{morpho}), is of order 50 to 100 $\mu$m.  Pozuelos et al.~(2015) described different Finson-Probstein models in which the size dependence of the velocity is varied.  They reported that $V \propto a^{-1/8}$ gives a better fit to the morphology than $V \propto a^{-1/2}$ or $V \propto a^{-1/20}$, but the effects are subtle and the significance of the difference is unclear.   Regardless, unless particle ejection from the nucleus is highly collimated in the orbital plane, a possibility which would seem to be physically improbable, then we may conclude that $V_{\perp}$ is only a few meters per second, and that it provides a measure of the total velocity of ejection of the dust.  

The low dust velocity accounts for the non-detection of coma in 313P.  A particle ejected towards the Sun at speed $V$ has a turn-around distance 

\begin{equation}
s = \frac{V^2}{2 \beta g_{\odot}}.
\end{equation}

\noindent Substituting $V$ = 2.5 m s$^{-1}$, $\beta$ = 0.01 to 0.02 and $g_{\odot}$ = 10$^{-3}$ m s$^{-2}$ gives $s$ = 160 to 310 km, corresponding to only 0.1\arcsec~to 0.2\arcsec~at 2 AU.  While technically resolvable with HST, this near-nucleus region of the surface brightness profile is dominated by the point-spread function of the vastly brighter nucleus (e.g.~see Figures \ref{sb_oct14} and \ref{sb_mar05}), making any coma difficult to detect.

We conclude that the dust is ejected from 313P at a speed two orders of magnitude smaller than the sound speed ($\sim$400 m s$^{-1}$) in sublimated gas at $\sim$3 AU, but slightly larger than the likely $V_e$ = 0.5 m s$^{-1}$ gravitational escape speed from the nucleus.   Very small launch speeds have been found in other active asteroids, most notably in the probable ice-sublimator 133P/Elst-Pizarro (Hsieh et al.~2004, Jewitt et al.~2014b).  Low terminal velocities can result from weak gas flow, as expected for ice at distances $r_H \sim$ 3 AU. Low terminal velocities can also be produced by the small size of the sublimating ice patch, which reduces the acceleration length over which gas drag can act (Jewitt et al.~2014b).

\label{spectra}

\section{DISCUSSION}

From Table (\ref{absolute}), it may be seen that the amount of  dust in 313P (represented by $C_d = C_e(6000) - C_n$) decreases from October to March by a factor of $\sim$5.  During this time, the heliocentric distance increased by only $\sim$8\%, from 2.4 to 2.6 AU (Table \ref{geometry}) and the sub-solar equilibrium temperature (which scales as $R^{-1/2}$) fell by only $\sim$4\%.  The implied  distance dependence (dust cross-section $\propto R^{-20}$) is unreasonably steep, and we conclude that the fading is  not caused by the changing distance.  The data in Figure (\ref{Hv_vs_DOY}) show that the coma absolute magnitudes have e-folding timescales $\sim$75 days, regardless of the radius of the photometry annulus used to measure the coma.   What determines the long timescale of the fading of 313P?

\subsection{Radiation Pressure Sweeping}
 Radiation pressure is capable of sweeping particles from the vicinity of the nucleus, but a simple calculation shows that the sweeping timescales are short compared to the $\sim$75 day fading timescale.  To see this, we note that the dominant particles as judged from syndyne/synchrone analyses have radiation pressure acceleration parameter $\beta \sim$ 0.01 to 0.02 (corresponding to particles of $\sim$50 to 100 $\mu$m size).  Neglecting the initial velocity of ejection, the time taken by solar radiation pressure to accelerate a dust particle over a distance $L$, is given by 

\begin{equation}
\tau = \left(\frac{2L}{\beta g_{\odot}}\right)^{1/2},
\label{time}
\end{equation}

\noindent  where $g_{\odot}$ is the local solar gravitational acceleration.  At 2.5 AU from the Sun, the solar gravitational acceleration is $g_{\odot}$ = 10$^{-3}$ m s$^{-2}$.  We consider, for example, dust in the 6000 km radius photometry aperture.  With $L$ = 6$\times$10$^6$ m, Equation (\ref{time}) gives $\tau \sim$ 8 to 11 $\times10^5$ s (9 to 13 days) and these are upper limits because of the neglect of the initial velocity.  This rapid clearing shows that radiation pressure is unlikely to produce the  decline in the coma cross-sections that occurs on timescales of $\sim$75 days.  Indeed, the persistence of the coma over five months requires that dust be continually released from the nucleus, evidently at a declining rate.  This photometry-based conclusion is consistent with the inference, made using Finson-Probstein dust dynamics models, that the mass loss occurred over a protracted period (Jewitt et al.~2015, Hsieh et al.~2015, Hui and Jewitt 2015, Pozuelos et al.~2015) and with an origin by sublimation.

Given that the dust is replenished on the timescale given by Equation (\ref{time}), we estimate the dust mass loss rate from the coma using

\begin{equation}
\frac{dM_d}{dt} = \frac{\rho \overline{a} C_d}{\tau}
\label{dmbdt}
\end{equation}

\noindent where $\rho = 10^3$ kg m$^{-3}$ is again the assumed bulk density of ejected grains, $\overline{a}$ is their effective radius, $C_d$ is the scattering cross section in the coma (Table \ref{absolute}) and $\tau$ is from Equation (\ref{time}).  We use $\overline{a}$ = 50 to 100 $\mu$m, as found from the syndyne/synchrone models.  Substitution into Equation (\ref{dmbdt}) gives values $dM_d/dt$ = 0.6 to 0.9 kg s$^{-1}$ in 2014 October declining to  $\sim$0.2 to 0.3 kg s$^{-1}$ by 2015 March.  These estimates compare with an upper limit to the gas production rate $\le$ 1.8 kg s$^{-1}$ based on the non-detection of emission lines in a spectrum taken on UT 2014 October 22 (Jewitt et al.~2015).  Peak dust production rates were independently estimated by Pozuelos et al.~(2015) at $dM_d/dt$ = 0.2 to 0.8 kg s$^{-1}$ (UT 2014 September 21).  Given the many uncertainties in the dust and model parameters, we consider these values in good agreement.

\subsection{Pit Source}
Instead, the long timescale of the coma fading leads us to consider illumination effects, of which we distinguish three types.  First, the orbit of 313P is eccentric, causing the heliocentric distance to vary from 2.391 AU at perihelion to 3.921 AU at aphelion.  This gives a factor $\sim$2.7 in solar insolation, which might be expected to have a measurable impact on the rates of sublimation at perihelion vs.~aphelion.  However, Table (\ref{geometry}) shows that the change in the heliocentric distance in any 75 day interval is negligible and, therefore, a decrease in the insolation is unlikely to be responsible for the observed fading.  Second, a seasonal variation of the solar insolation at any point on the surface of 313P will result from non-zero obliquity of the spin, being largest for obliquity = 90\degr, when the Sun can cross both the projected equator and the poles of 313P at different points in the orbit.  Third, the pattern of local shadows on the surface will be modulated by the position of 313P in its orbit.  Since the sublimating area on 313P is very small (Jewitt et al.~2015) there is reason to expect that local shadows may play an important role in modulating the mass loss.

We consider a simple model to attempt to capture the second and third of these effects.  In this model, the instantaneous equilibrium sublimation rate, $F_s$ (kg m$^{-2}$ s$^{-1}$), is calculated from the energy balance equation

\begin{equation}
\frac{F_{\odot} (1 - A) }{r_H^2} \cos(i(t)) =  \epsilon \sigma T^4  + L(T) F_s(t)
\label{energy}
\end{equation}

\noindent in which $F_{\odot}$ = 1360 W m$^{-2}$ is the Solar constant, $A$ is the Bond albedo, $\epsilon$ is the emissivity, $\sigma$ is the Stefan-Boltzmann constant, $L(T)$ is the latent heat of sublimation for water ice at temperature, $T$, and $F_s$ is the equilibrium sublimation flux (kg m$^{-2}$ s$^{-1}$).    We took $A$ = 0.05, $\epsilon$ = 0.9, $\sigma$ = 5.67$\times$10$^{-8}$ W m$^{-2}$ K$^{-4}$, while $L(T)$ was obtained from Washburn (1926) and all calculations were performed for $r_H$ = 2.5 AU, representative of the heliocentric distance of 313P (Table \ref{geometry}).  

The angle $i(t)$ in Equation (\ref{energy}) is the zenith angle of the Sun as viewed from the sublimating surface.  Angle $i$ (which must be  $\le 90\degr$ for Equation (\ref{energy}) to hold)  is determined by the instantaneous orientation of the nucleus spin vector relative to the Sun direction, by $\psi$, the latitude of the sublimating patch, by the local slope of the surface and by the time of day.  Neither the shape nor the spin vector of 313P are well known.  In the following, we assume that the  obliquity is 90\degr~in order to maximize the seasonal effects.  In this geometry, and assuming a spherical nucleus with a spin vector perpendicular to the Sun-nucleus line,  the local solar zenith angle may be written as $\cos(i(t)) = \cos(\psi) \cos(\omega t)$, where $\omega = 2\pi/P$ is the angular frequency of the rotation, $P$ is the rotation period and $t$ is the time.  We arbitrarily take $P$ = 4 hr, although the period is unimportant for the relative effects described here.

\begin{figure}
\epsscale{0.85}
\begin{center}
\plotone{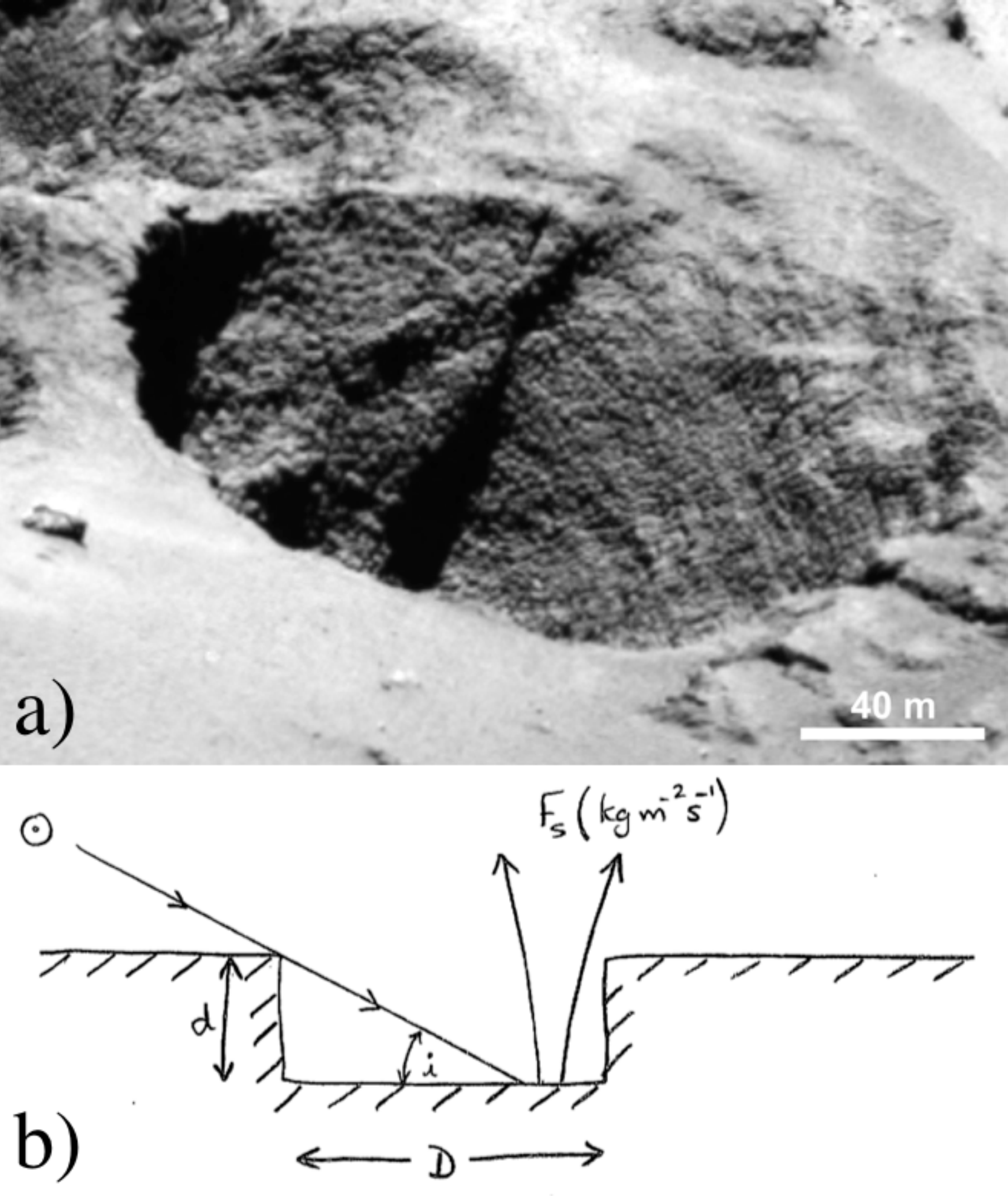}
\caption{a)~Oblique image of a deep sublimation pit (depth to diameter ratio $\sim$1) on the nucleus of Jupiter-family comet 67P/Churyumov-Gerasimenko recorded from the ESA Rosetta spacecraft (credit:~ESA/Rosetta/MPS/OSIRIS) b) sketch defining parameters of the  pit model in Equation (\ref{DeltaM}).   \label{pit}
} 
\end{center} 
\end{figure}

\begin{figure}
\epsscale{1.0}
\begin{center}
\plotone{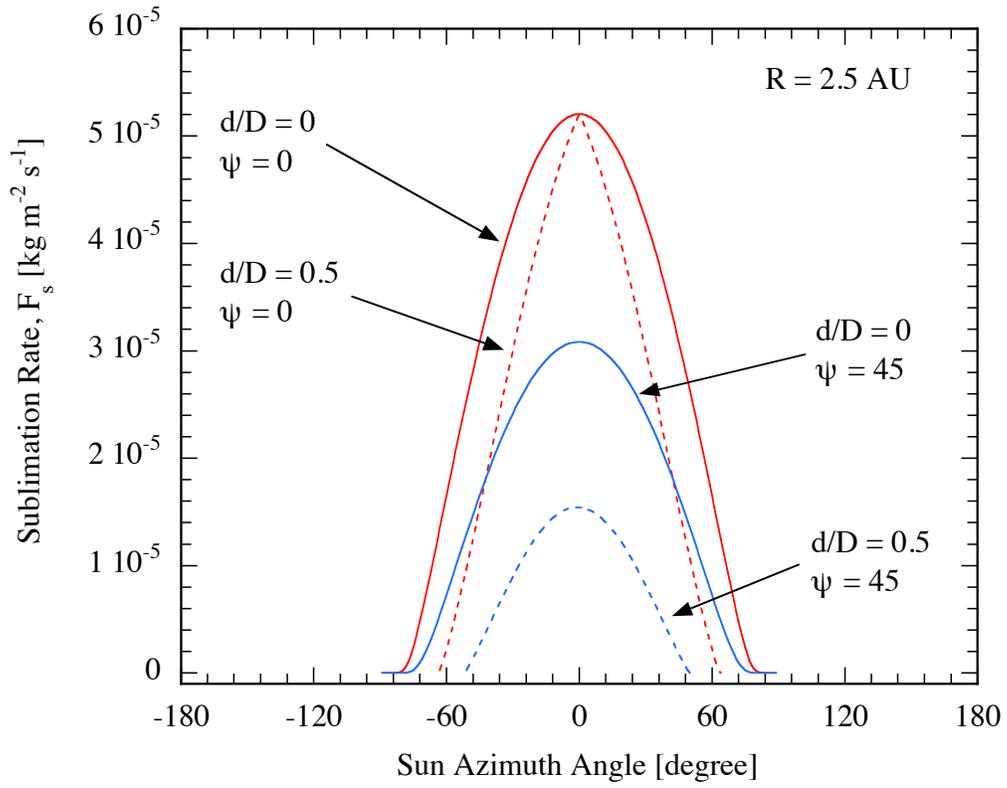}
\caption{Sublimation from ice exposed at two latitudes ($\psi$ = 0\degr~ and 45\degr) and two pit depths ($d/D$ = 0 and 0.5), as marked, as a function of the instantaneous solar azimuth angle.     \label{cosine_plot}
} 
\end{center} 
\end{figure}

Solutions to Equation (\ref{energy}) are plotted in Figure (\ref{cosine_plot}) for the diurnal variation of the ice sublimation rate on the rotational equator ($\psi$ = 0\degr) and at $\psi$ = 45\degr, as solid red and blue curves, respectively.  The figure shows that the largest sublimation rates at 2.5 AU are $F_s \sim 5\times10^{-5}$ kg m$^{-2}$ s$^{-1}$ on the equator, falling to about 60\% of this value at 45\degr~latitude.  As expected, sublimation at 45\degr~is limited in duration as well as in rate relative to the equatorial case.  Continued mass loss from a region of the surface at the  rate $F_s$ (kg m$^{-2}$ s$^{-1}$) will  cause the sublimating surface to sink relative to the surrounding non-sublimating surface at rate $d\ell/dt = F_s/ \rho$, where $\rho$ is the bulk density of the surface layers.  With $\rho$ = 10$^3$ kg m$^{-3}$, the surface recession rate is $d\ell/dt = 5 \times 10^{-8}$ m s$^{-1}$, or 4 mm day$^{-1}$.  Progressive sublimation will create a pit whose floor remains partly self-shadowed, leading to  variations of the sublimation rate on both diurnal and seasonal timescales.  Sublimation pits have long been discussed in the context of active comets (e.g.~Keller et al.~1994), although their formation mechanism is uncertain (Thomas et al.~2013). They have recently been imaged in spectacular detail on the nucleus of comet 67P/Churyumov-Gerasimenko (Figure \ref{pit}a).

We explored a simple model of self-shadowing, in which the pit is rectangular in cross-section with a depth-to-diameter ratio $d/D$ (see Figure \ref{pit}b).  We assume that sublimation  proceeds from the floor of the pit, although there is some suggestion that at least one pit on 67P sublimates through its walls (Vincent et al.~2015).  The detailed geometry is not  crucial to the argument that follows; what matters most is that heating of the ice is topographically obstructed, and this depends mainly on the ratio of the depth (or height) to the horizontal scale of the sublimating patch.  For the assumed pit source, the fraction of the floor illuminated by the Sun is 

\begin{equation}
f(t) \sim 1-\frac{d}{D\tan (i(t))}
\label{fraction}
\end{equation}

\noindent provided $i \ge i_c = \tan^{-1}(d/D)$,  and $f(t) = 0$ otherwise.   Equations  (\ref{energy}) and (\ref{fraction})  were used to compute

\begin{equation}
\Delta M = \int_0^P f(t) F_s(t) dt
\label{DeltaM}
\end{equation}

\noindent where the integral is taken over one rotation of the nucleus.  Equation (\ref{DeltaM}) gives the total mass lost  in one nucleus rotation per unit area by equilibrium sublimation from a surface element at latitude $\psi$.    

\begin{figure}
\epsscale{0.85}
\begin{center}
\plotone{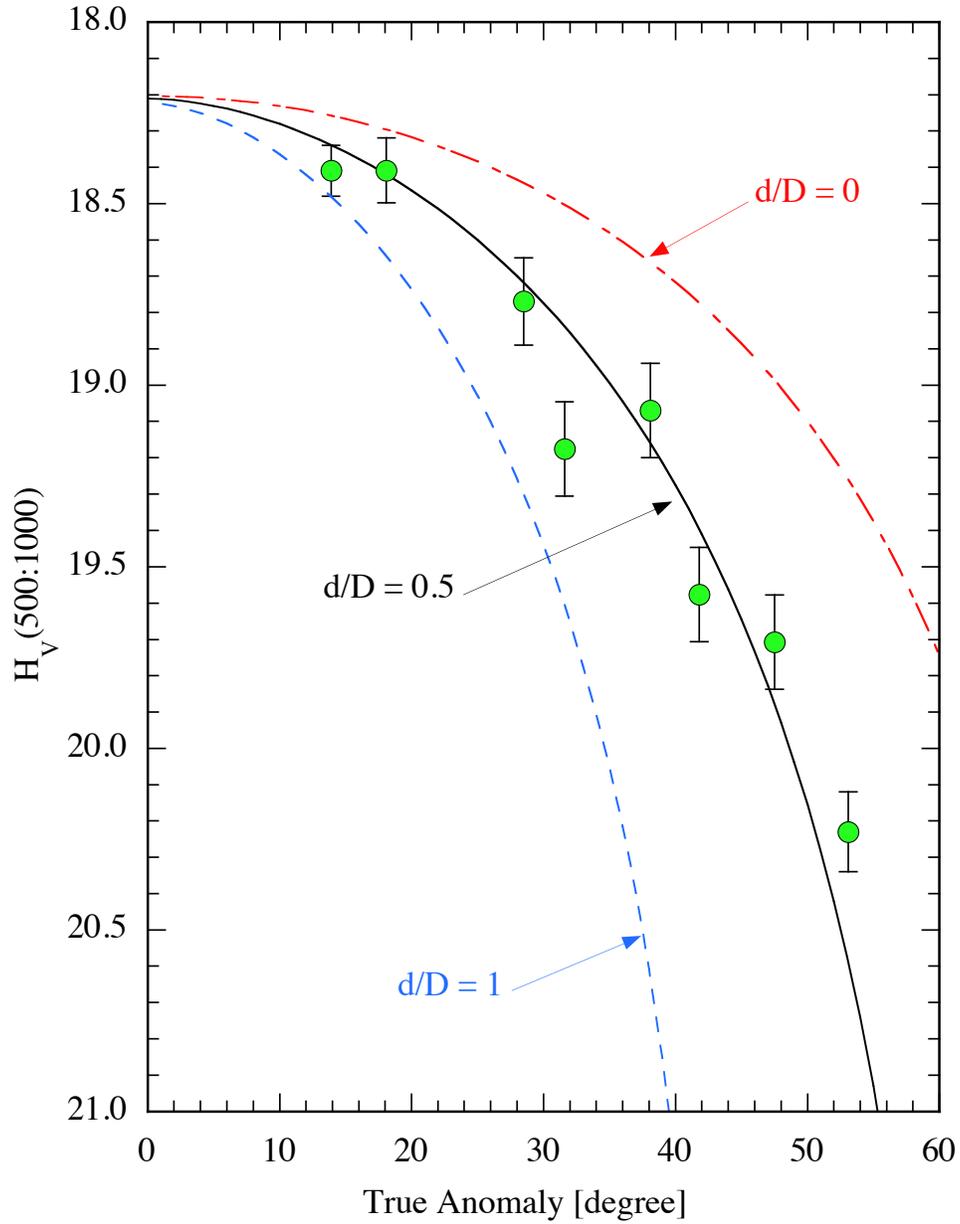}
\caption{The absolute magnitude, $H_V(500$:$1000)$ (green circles) as a function of the true anomaly.  Three pit source sublimation models having depth to diameter ratios $d/D$ = 0.0 (red dash-dot line), 0.5 (black solid line) and 1.0 (blue dashed line) show the effect of self-shadowing for an equatorial source on a nucleus with 90\degr~obliquity.     \label{true_anom_fit}
} 
\end{center} 
\end{figure}

Figure (\ref{true_anom_fit}) shows a comparison between the photometry and the model results from Equation (\ref{DeltaM}).  For clarity of presentation we show only measurements from the 500$:$1000 km annulus, but other annular photometry measurements give consistent results, since they show similar rates of fading.  To make the figure we have assumed that the sublimated mass from Equation (\ref{DeltaM}) and the scattering cross-section (and hence the measured brightness) are proportional, $\Delta M \propto C_e$.    We find that models with $d/D$ = 0 (i.e.~surface sublimation) and $d/D$ = 1 are strongly inconsistent with the measured fading rates because they predict brightness variations with mean anomaly that are, respectively, too flat and too steep  to match the observations.  On the other hand, models with $d/D \sim$ 1/2 better match the rate of decline in the brightness of 313P by virtue of the inclusion of self-shadowing in the pit.  

The curves in Figure (\ref{true_anom_fit}) are non-unique, and the model behind them is highly simplified.  For instance, we are forced to assume the  obliquity in the absence of constraining evidence and we assumed that the nucleus of 313P is  spherical when such a small body is unlikely to be so.  A different obliquity and a different nucleus body-shape would give different results. We have also neglected any possibility of an insulating dust mantle on the sublimating ice at the bottom of the pit.  This is likely a serious omission, in that pit sublimation may be stifled both by self-shadowing and by the accumulation of debris (see Guilbert-Lepoutre et al.~2015 for a modern discussion of these effects).  Nevertheless, our model is sufficient to show that seasonal effects caused by non-zero obliquity and local shadowing in a pit source can easily match the observed fading rate, and give a physically plausible pit geometry.  This conclusion can be extended to other active asteroids, for instance 133P/Elst-Pizarro (Hsieh et al. 2004), in which activity is known to be restricted to a substantial range of true anomalies.   

For an equatorially located pit with $d/D$ = 1/2, Figure (\ref{cosine_plot}) indicates a rotationally-averaged mass loss rate $F_s \sim$ 1$\times$10$^{-5}$ kg m$^{-2}$ s$^{-1}$.  To compare with the estimates of the dust mass production rate (near 1 kg s$^{-1}$ at peak), we need to know the ratio of dust to gas production rates, $f_{dg}$. Measurements of active comets generally give values $f_{dg} >$ 1 (this is possible because the escaping gas  travels much faster than the ejected dust, maintaining momentum equipartition between the two components).  For example, observations of comet 2P/Encke give 10 $\le f_{dg} \le$ 30 (Reach et al.~2000).   The area of exposed ice needed to supply a dust mass loss rate $dM_d/dt$ is

\begin{equation}
\pi r_s^2 = \left(\frac{1}{f_{dg} F_s} \right) \frac{dM_d}{dt}.
\end{equation}

\noindent With $dM_d/dt$ = 1 kg s$^{-1}$, as inferred above, $F_s$ = 1$\times$10$^{-5}$ kg m$^{-2}$ s$^{-1}$ and conservatively taking $f_{dg}$ = 10, we find $\pi r_s^2$ = 10$^4$ m$^2$, and $r_s$ = 56 m. This sublimating area corresponds to $\sim$0.2\% of the surface of a 700 m radius spherical nucleus.  It is possible that the source is a single region of radius and depth $d = r_s \sim$ 56 m, perhaps formed by a small impact.  Such a pit would take a time $d\rho/(F_s) \sim$ 6$\times$10$^9$ s (200 years) to grow.   We think it more likely that the sublimating area consists of a number of pits, each less deep and younger than this estimate.  For example,  $N$ equal-size pits of radius and depth $\sim$56/$N^{1/2}$  would satisfy the observations.  With $N$ = 100, each pit would be $\sim$6 m deep and have an excavation time $\sim$20 years.  Unfortunately, we possess no observational constraint on $N$.  

Low dust velocities in 133P were interpreted as evidence for sublimation from patches of limited size, because  the acceleration length and the terminal velocity  for gas-entrained dust particles scale with the physical size of the sublimating source region (Jewitt et al.~2014b).  Dust in 313P is ejected with small velocities comparable to those in 133P (c.f.~Figure \ref{FWHM}), and sublimation from a set of small pits may again be responsible.  It may be natural to expect that an exposed ice surface should develop into a set of pits in response to fallback mantling, the presence of surface irregularities such as boulders and recession of the sublimating surface beneath the physical surface of the adjacent nucleus.

\subsection{Variation Around the Orbit}
\begin{figure}
\epsscale{0.95}
\begin{center}
\plotone{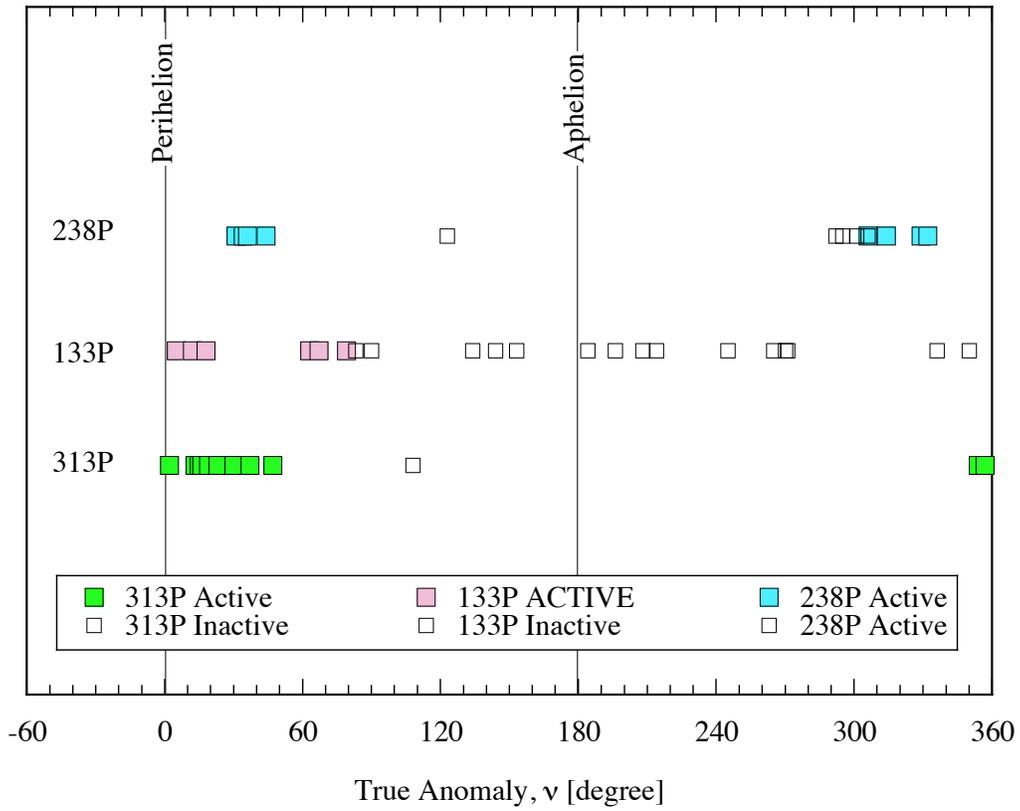}
\caption{Activity state of the three repetitively active asteroids as a function of the true anomaly at the time of observation. Detections of activity are marked using filled squares while non-detections are marked with unfilled squares.  \label{true_nu}
} 
\end{center} 
\end{figure}

Seasonal effects in all three  of the active asteroids that have displayed repetitive activity in different orbits are plotted in Figure (\ref{true_nu}). These are the strongest candidates for being driven by the sublimation of ice.  The data were compiled for 133P from Hsieh et al.~(2010), for 238P from Hsieh et al.~(2011) and for 313P from Jewitt et al.~(2015), Hsieh et al.~(2015), Hui and Jewitt (2015) and from the present work.  There is an absence of activity within the  true anomaly range $\nu \sim 180 \pm$90\degr~for each object.  Activity is observed to about $\pm$60\degr~in 238P.  Prototype 133P shows activity mainly after perihelion, up to $\nu \sim$ 80\degr, suggesting the action of a thermal lag (Hsieh et al.~2010).  The variation of activity around the orbit shown in Figure (\ref{true_nu}) is consistent with seasonal modulation  on all three objects, as expected if sublimating ice is responsible.  However, it should be pointed out that the full range of true anomaly is incompletely sampled, especially for 238P and 313P, and it is  possible that activity occurs over a wider fraction of each orbit than existing observations reveal.  In addition, it is likely that observational bias favors the detection of objects near perihelion because objects are brighter there than at aphelion.    For example,  typical perihelion and aphelion distances of the three above objects are $\sim$2.5 AU and 4.0 AU, respectively.  Observed as point sources at opposition (corresponding to geocentric distances $\sim$1.5 AU and 3.0 AU, respectively), the inverse square law predicts the  objects to be fainter at aphelion than perihelion by $\sim$2.5 magnitudes.  In addition, the ratio of the  equilibrium sublimation rates at perihelion and aphelion is an order of magnitude or more, constituting an observationally formidable obstacle to the detection of activity at large distances.  More work is needed to buttress the apparent concentration of activity in these bodies near perihelion.

\subsection{Timescales}
The inferred mass loss rate $dM_d/dt \sim$ 1 kg s$^{-1}$, if sustained over 5 months, would correspond to a total mass lost from 313P of $\Delta M_d \sim$10$^7$ kg per $P_K$ = 5.6 year orbit.  
For comparison, the mass of the nucleus, represented as a sphere of radius $r_n$ = 700 m and density $\rho$ = 10$^3$ kg m$^{-3}$ is $M_n \sim$ 10$^{12}$ kg.  Thus, the fractional mass loss per orbit is $\Delta M_d/ M_n \sim$ 10$^{-5}$ and 313P could sustain continuous activity at the present rate for a time $P_K M_n /\Delta M_d \sim$ 6$\times$10$^5$ yr.  The age of 313P is unknown but is likely to be much greater.   Hsieh et al.~(2015) report a dynamical association with the 160$\pm$35 Myr old Lixiahua asteroid family (c.f.~Novakovic et al.~2010) which, if real, would suggest that 313P is about 100 times older than its sublimation age. The approximate collisional destruction time for a  $r_n$ = 700 m radius asteroid is even older at $\sim$10$^9$ yr (Bottke et al.~2005).  The survival of ice in 313P for 10$^8$ and 10$^9$ yr periods can be simply explained by the presence of a refractory  mantle thick enough to stifle sublimation of near-surface ice (Schorghofer 2008).   Such a mantle would block sublimation until penetrated, perhaps by impact or by surface instability (Hsieh and Jewitt 2006). The duty cycle for activity (i.e.~the fraction of the time for which the body is active) need only be $\lesssim 10^{-2}$  for ice to survive for the age of the Lixiahua family or $\lesssim 10^{-3}$ to survive for the collisional lifetime of the object, respectively.  A duty cycle of $\lesssim 10^{-2}$ or $\lesssim 10^{-3}$ implies that, for every observed active case like that of 313P, there are $\gtrsim 10^2$ or $\gtrsim 10^3$ similar but dormant, ice-containing bodies in the asteroid belt.

\clearpage

\section{SUMMARY}

313P/Gibbs is an active asteroid ejecting dust near perihelion in two different orbits, suggesting an origin by the intermittent sublimation of near-surface ice.  We have used the Hubble Space Telescope to study this object in detail, in order to better constrain the nature of its activity.  We find that

\begin{enumerate}

\item The  nucleus radius, estimated from convolution models of the surface brightness profile, lies in the range 0.6 to 0.8 km (geometric albedo 0.05 assumed).

\item The dust distribution  is consistent with the continued ejection of large particles (radiation pressure parameter $\beta$ = 0.01 to 0.02, corresponding to particle radii $\sim$50 to 100 $\mu$m) in observations extending from 2014 October to 2015 March.

\item In-plane observations show that these particles are ejected slowly, with characteristic velocities normal to the orbit plane of $\sim$2.5 m s$^{-1}$.  

\item The peak mass loss rate in dust is of order 1 kg s$^{-1}$, decreasing on an e-folding time $\sim$75 day.  The area of exposed ice needed to supply this rate is $\sim$10$^4$ m$^2$, corresponding to only $\sim$0.2\% of the nucleus surface.

\item The secular fading is too slow to be  caused by radiation pressure sweeping of dust particles from the coma but too fast to be related to the (marginally) increasing heliocentric distance.  Instead, we show by a simple model that the fading timescale is consistent with protracted sublimation of ice from a topographically shadowed region, for example in a pit  having depth-to-diameter ratio $\sim$0.5.  Such pits are a natural product of localized ice sublimation.

\item The ejected dust mass is $\sim$10$^7$ kg per orbit, approximately 10$^{-5}$ of the 10$^{12}$ kg nucleus mass.  If comparably active in every orbit, the mass-loss lifetime is confined to $\sim$0.6 Myr, far smaller than the $\sim$160$\pm$35 Myr age of the Lixiahua family of which 313P is a likely member, and smaller than the $\sim$1 Gyr lifetime to collisions.  To reconcile these timescales requires that the duty cycle for sublimation-driven mass-loss be $\lesssim$10$^{-2}$ to $\lesssim$10$^{-3}$, respectively.

\end{enumerate}

\acknowledgments
Based in part on observations made with the NASA/ESA \emph{Hubble Space Telescope,} with data obtained via the Space Telescope Science Institute (STSCI).  Support for program 14040  was provided by NASA through a grant from STSCI, operated by AURA, Inc., under contract NAS 5-26555.  We thank Linda Dressel, Alison Vick and other members of the STScI ground system team for their expert help and the anonymous referee for comments.

\clearpage

\clearpage

\begin{deluxetable}{lclccccccr}
\tablecaption{Observing Geometry 
\label{geometry}}
\tablewidth{0pt}
\tablehead{ \colhead{UT Date and Time} & DOY\tablenotemark{a}   & $\Delta T_p$\tablenotemark{b} & $\nu$\tablenotemark{c} & \colhead{$R$\tablenotemark{d}}  & \colhead{$\Delta$\tablenotemark{e}} & \colhead{$\alpha$\tablenotemark{f}}   & \colhead{$\theta_{\odot}$\tablenotemark{g}} &   \colhead{$\theta_{-v}$\tablenotemark{h}}  & \colhead{$\delta_{\oplus}$\tablenotemark{i}}   }
\startdata

2014 Oct 14 13:12 - 13:48 & 287 & 47 & 14 &  2.405 & 1.451 & 8.9 & 10.5 & 247.9 & 7.5 \\
2014 Oct 28 21:25 - 23:16 & 301 & 61 & 18 &   2.415 & 1.522 & 13.0 & 37.1 & 248.8 & 6.8 \\

2014 Dec 01 04:36 - 05:13 & 335 & 95 & 28 & 2.446 & 1.835 & 21.0 & 58.8 & 249.2 & 3.6 \\
2014 Dec 15 07:36 - 08:12 & 349 & 109 & 32 &  2.463 &  2.011 & 22.6 & 62.9 & 248.7 & 2.1 \\
2015 Jan 07 16:20 - 18:08 & 372 & 132 & 38 & 2.495  & 2.326 & 23.2 & 67.7 & 247.6 & -0.0 \\
2015 Jan 21 03:07 - 03:43 & 386  & 146 & 42 & 2.516 & 2.513 & 22.6 & 69.9 & 247.1 & -1.0 \\
2015 Feb 11 19:37 - 20:14 & 407  & 167& 47  & 2.553 & 2.809 & 20.5 & 73.3 & 246.8 & -2.2 \\
2015 Mar 05 14:00 - 15:48 & 429 & 189 & 53 & 2.593 & 3.085 & 17.5 & 76.9 & 247.1 & -3.0 \\

\enddata


\tablenotetext{a}{Day of Year, UT 2014 January 01 = 1}
\tablenotetext{b}{Number of days past perihelion (UT 2014 August 28 = DOY 240)}
\tablenotetext{c}{True anomaly, in degrees}
\tablenotetext{d}{Heliocentric distance, in AU}
\tablenotetext{e}{Geocentric distance, in AU}
\tablenotetext{f}{Phase angle, in degrees}
\tablenotetext{g}{Position angle of the projected anti-Solar direction, in degrees}
\tablenotetext{h}{Position angle of the projected negative heliocentric velocity vector, in degrees}
\tablenotetext{i}{Angle of Earth above the orbital plane, in degrees}

\end{deluxetable}

%
\begin{deluxetable}{lcccr}
\tabletypesize{\scriptsize}
\tablecaption{Photometry with Fixed Angle Apertures
\label{photometry}}
\tablewidth{0pt}
\tablehead{
\colhead{UT Date}    & \colhead{$\Phi$\tablenotemark{a}   }   & \colhead{$m_{V}\tablenotemark{b}$} & \colhead{$H_V$\tablenotemark{c}} & \colhead{$C_e [km^2]\tablenotemark{d}$} 
}

\startdata
Oct 14 & 0.2  & 20.62$\pm$0.01 & 17.3$\pm$0.07  & 3.3$\pm$0.2 \\
Oct 14 & 1.0  & 20.02$\pm$0.01 & 16.7$\pm$0.07  & 5.8$\pm$0.4 \\
Oct 14 & 4.0  & 19.60$\pm$0.01 & 16.3$\pm$0.07  & 8.5$\pm$0.6 \\
Oct 14 & 6.0  & 19.40$\pm$0.01 & 16.1$\pm$0.07  & 10.2$\pm$0.7 \\\\

Oct 28 & 0.2  & 20.97$\pm$0.01 & 17.4$\pm$0.09  & 3.1$\pm$0.3 \\
Oct 28 & 1.0  & 20.36$\pm$0.01 & 16.8$\pm$0.09  & 5.4$\pm$0.5 \\
Oct 28 & 4.0  & 19.81$\pm$0.01 & 16.2$\pm$0.09  & 9.0$\pm$0.8 \\
Oct 28 & 6.0  & 19.72$\pm$0.01 & 16.1$\pm$0.09  & 9.8$\pm$0.9 \\\\

Dec 01 & 0.2 & 21.83$\pm$0.01 &  17.54$\pm$0.12  &  2.7$\pm$0.3 \\
Dec 01 & 1.0 & 21.27$\pm$0.01 &  16.98$\pm$0.12   &  4.5$\pm$0.5   \\
Dec 01 & 4.0 & 20.63$\pm$0.01  & 16.34$\pm$0.12   &  8.0$\pm$1.0  \\
Dec 01 & 6.0& 20.43$\pm$0.01 &  16.14$\pm$0.12   &   9.7$\pm$1.2 \\\\

Dec 15 & 0.2 & 22.17$\pm$0.01 & 17.62$\pm$0.13    &   2.5$\pm$0.3  \\
Dec 15 & 1.0 & 21.66$\pm$0.01 &   17.11$\pm$0.13  &  3.9$\pm$0.5  \\
Dec 15 & 4.0 & 21.01$\pm$0.01 &  16.46$\pm$0.13    &  7.2$\pm$0.9   \\
Dec 15 & 6.0 & 20.85$\pm$0.01 &    16.30$\pm$0.13 &  8.3$\pm$1.1  \\\\

Jan 07 & 0.2 & 22.67$\pm$0.01 & 18.12$\pm$0.13    &  1.6$\pm$0.2  \\
Jan 07 & 1.0  & 22.08$\pm$0.01 &   17.53$\pm$0.13  &  2.7$\pm$0.3  \\
Jan 07 & 4.0 & 21.28$\pm$0.01 &  16.73$\pm$0.13   &   5.6$\pm$0.7  \\
Jan 07 & 6.0 & 21.01$\pm$0.01 &  16.46$\pm$0.13   & 7.2$\pm$0.9   \\\\

Jan 21 & 0.2 & 22.96$\pm$0.01 & 17.87$\pm$0.13 & 2.0$\pm$0.3  \\
Jan 21 & 1.0 & 22.48$\pm$0.01 & 17.39$\pm$0.13 & 3.1$\pm$0.4 \\
Jan 21 & 4.0 & 21.78$\pm$0.01 & 16.69$\pm$0.13 & 5.8$\pm$0.8 \\
Jan 21 & 6.0 & 21.74$\pm$0.01 & 16.65$\pm$0.13 & 6.0$\pm$0.8 \\\\

Feb 11 & 0.2 & 23.05$\pm$0.01 & 17.76$\pm$0.13  & 2.2$\pm$0.3 \\
Feb 11 & 1.0 & 22.68$\pm$0.01 & 17.39$\pm$0.13 & 3.1$\pm$0.4 \\
Feb 11 & 4.0 & 22.18$\pm$0.01 & 16.89$\pm$0.13 & 4.8$\pm$0.6 \\
Feb 11 & 6.0 & 22.06$\pm$0.01 & 16.77$\pm$0.13 & 5.4$\pm$0.7 \\\\

Mar 05 & 0.2 & 23.27$\pm$0.01 & 17.83$\pm$0.11 & 2.0$\pm$0.2 \\
Mar 05 & 1.0 & 22.57$\pm$0.01 & 17.53$\pm$0.11 & 2.7$\pm$0.2\\
Mar 05 & 4.0 & 22.63$\pm$0.01 & 17.19$\pm$0.11 & 3.7$\pm$0.4 \\
March 05 & 6.0 & 22.60$\pm$0.01 & 17.16$\pm$0.11 & 3.8$\pm$0.4 \\
   
\enddata


\tablenotetext{a}{Angular radius of photometry aperture, in arcsec}
\tablenotetext{b}{Apparent V magnitude}
\tablenotetext{c}{Absolute magnitude computed from Equation (\ref{absolute}).  }
\tablenotetext{d}{Effective scattering cross-section computed from Equation (\ref{area}), km$^2$}

\end{deluxetable}

\clearpage 

\begin{deluxetable}{lccccccc}
\tabletypesize{\scriptsize}
\tablecaption{Photometry with Fixed Linear Apertures
\label{absolute}}
\tablewidth{0pt}
\tablehead{
\colhead{UT Date}    & \colhead{Quantity\tablenotemark{a}} & \colhead{500 km }   & \colhead{1000 km} & \colhead{2000 km} & \colhead{4000 km} & \colhead{6000 km}
}

\startdata

Oct 14 	&	V 			& 20.30	 	& 20.04	 	& 19.77	 		& 19.53	 		& 19.44 \\
Oct 14      &	H$_V$ 		&16.99		& 16.73	 	& 16.46	 		& 16.22	 		& 16.13 \\
Oct 14 	& $C_e$ 			& 4.4 		& 5.6			& 7.2				& 9.0				& 9.7 \\\\
	
Oct 28   	 &	V 			& 20.67	 	& 20.39		& 20.11		 	& 19.77		 	& 19.67 \\
Oct 28 	 &	H$_V$ 		& 17.08	 	& 16.80	 	& 16.52		 	& 16.18		 	& 16.08 \\
Oct 28 	& $C_e$ 			& 4.1  		& 5.3	 		& 6.8	 			& 9.3	 			& 10.2 \\\\

Dec 01 	 &	V 			& 21.64	 	& 21.38	 	& 21.10		 	& 20.79		 	& 20.59 \\ 
Dec 01 	 &	H$_V$ 		& 17.35	 	& 17.09	 	& 16.81		 	& 16.50		 	& 16.30 \\
Dec 01 	& $C_e$ 			& 3.2	  		& 4.0	 		& 5.2	 			& 6.9 			& 	8.3 \\\\
	
Dec 15 	&	V  			& 21.99	 	& 21.79		& 21.55		 	& 21.25		 	& 21.05 \\
Dec 15 	 &	H$_V$ 		& 17.44	 	& 17.24	 	& 17.00		 	& 16.70		 	& 16.50 \\
Dec 15	& $C_e$ 			& 2.9 	 	& 3.5	 		& 4.4	 			& 5.8	 			& 6.9 \\\\

Jan 07 	&	V  			& 22.57	 	& 22.31	 	& 22.03		 	& 21.67		 	& 21.45 \\	
Jan 07 	&	H$_V$ 		& 17.65	 	& 17.39		& 17.11		 	& 16.75		 	& 16.53 \\
Jan 07 	& $C_e$ 			& 2.4	  		& 3.1	 		& 4.0	 			& 5.5	 			& 6.7 \\\\

Jan 21 	&	V  			& 22.86	 	& 22.67		 & 22.46		 	& 22.10		 	& 21.94 \\
Jan 21 	 &	H$_V$ 		& 17.78	 	& 17.59		 & 17.38		 	& 17.11		 	& 16.86 \\
Jan 21 	& $C_e$ 			& 2.1 	 	& 2.5	 		& 3.1	 			& 4.0	 			& 5.0 \\\\

Feb 11 	&	V  			& 23.01	 	& 22.85		 & 22.70		 	& 22.52		 	& 22.40 \\
Feb 11 	&	H$_V$  		& 17.71	 	& 17.55		 & 17.40		 	& 17.22		 	& 17.10 \\
Feb 11 	& $C_e$ 			& 2.3  		& 2.6	 		& 3.0	 			& 3.6	 			& 4.0 \\\\

Mar 05 	&	V  			& 23.24	 	& 23.13		 & 23.01		 	& 22.85		 	& 22.79 \\
Mar 05 	 &	H$_V$ 		& 17.80	 	& 17.69		 & 17.57		 	& 17.41		 	& 17.35 \\
Mar 05	& $C_e$ 			& 2.1  		& 2.3	 		& 2.6	 			& 3.0	 			& 3.2 \\\\

\enddata

\tablenotetext{a}{V = apparent V magnitude, $H_V$ = absolute V magnitude, $C_e$ = effective scattering cross-section in km$^2$}

\end{deluxetable}

\clearpage 

\begin{deluxetable}{cccccc}
\tablecaption{Surface Brightness Fits
\label{profile_fit}}
\tablewidth{0pt}
\tablehead{
\colhead{UT Date}    & \colhead{$p$\tablenotemark{a}} & \colhead{$C_{0.2}$\tablenotemark{b}} & \colhead{$F$\tablenotemark{c}}     & \colhead{$C_n$\tablenotemark{d}} & \colhead{$r_n$\tablenotemark{e}} 
}

\startdata

Oct 14 	&	1.60$\pm$0.01 	& 3.3$\pm$0.3		& 0.66$\pm$0.10				& 2.2$\pm$0.4  			& 0.8$\pm$0.1\\
Oct 28 	&	1.60$\pm$0.01 	& 3.1$\pm$0.3		& 0.66$\pm$0.10 				& 2.0$\pm$0.4  			& 0.8$\pm$0.1 \\
Dec 01 	&	1.52$\pm$0.01 	& 2.7$\pm$0.3	 	& 0.80$\pm$0.10 				& 2.2$\pm$0.4 				&  0.8$\pm$0.1 \\
Dec 15 	&	1.77$\pm$0.01 	& 2.5$\pm$0.3		& 0.66$\pm$0.10	 	 		& 1.7$\pm$0.3 				& 0.7$\pm$0.1  \\
Jan 07 	&	1.42$\pm$0.01 	& 1.6$\pm$0.2		& 0.70$\pm$0.10	 	 			& 1.1$\pm$0.2 			& 0.6$\pm$0.1  \\
Jan 21 	&	1.65$\pm$0.01 	& 2.0$\pm$0.3		&0.64$\pm$0.10	 	& 1.3$\pm$0.3	 					& 0.6$\pm$0.1 \\
Feb 11\tablenotemark{f} 	&	-- 					& --	 	& --	 	& --	 			& --  \\
Mar  05 	&	2.08$\pm$0.10					& 2.2$\pm$0.2	& 0.80$\pm$0.10	 	& 1.8$\pm$0.3	 		& 0.7$\pm$0.1  \\

\enddata

\tablenotetext{a}{Surface brightness gradient index measured in the angular radius range 0.2\arcsec~$\le \theta \le$ 1.0\arcsec}
\tablenotetext{b}{Cross-section (km$^2$) inside a  0.2\arcsec~radius circle, from Table \ref{photometry}}
\tablenotetext{c}{Fraction of $C_{0.2}$ contributed by the nucleus, determined from the convolution model}
\tablenotetext{d}{Nucleus cross-section (km$^2$)}
\tablenotetext{e}{Effective nucleus radius, $r_n = (C_n/\pi)^{1/2}$, (km)}
\tablenotetext{f}{No reliable fit for $p$ was possible using data from this date}

\end{deluxetable}

\clearpage

%
%
%
%
%
%
%
%
%
%
%
%
%
%
%

\clearpage


\clearpage

\clearpage

\clearpage

\clearpage

\clearpage

\clearpage

\clearpage

\clearpage

\clearpage

\clearpage

\clearpage

\clearpage

\begin{thebibliography}{}
\bibitem[Bohren \& Huffman(1983)]{1983asls.book.....B} Bohren, C.~F., \& Huffman, D.~R.\ 1983, New York: Wiley
%
\bibitem[Bottke et al.(2005)]{2005Icar..179...63B} Bottke, W.~F., Durda, D.~D., Nesvorn{\'y}, D., et al.\ 2005, Linking the collisional history of the main asteroid belt to its dynamical excitation and depletion, \icarus, 179, 63 
%
\bibitem[Bowell et al.(1989)]{1989aste.conf..524B} Bowell, E., Hapke, B., Domingue, D., et al.\ 1989, Asteroids II, 524 
%
\bibitem[Cotto-Figueroa et al.(2015)]{2015ApJ...803...25C} Cotto-Figueroa, D., Statler, T.~S., Richardson, D.~C., \& Tanga, P.\ 2015, \apj, 803, 25 
%
%
%
%
%
\bibitem[Fern{\'a}ndez et al.(2013)]{2013Icar..226.1138F} Fern{\'a}ndez, Y.~R., Kelley, M.~S., Lamy, P.~L., et al.\ 2013, \icarus, 226, 1138 
%
\bibitem[Finson \& Probstein(1968)]{1968ApJ...154..327F} Finson, M.~J., \& Probstein, R.~F.\ 1968, \apj, 154, 327 
%
%
\bibitem[Gibbs 
\& Sato(2014)]{2014CBET.3991....1G} Gibbs, A.~R., \& Sato, H.\ 2014, Central Bureau Electronic Telegrams, 3991, 1 

%
\bibitem[Guilbert-Lepoutre et al.(2015)]{2015SSRv..tmp...23G} Guilbert-Lepoutre, A., Besse, S., Mousis, O., et al.\ 2015, \ssr, 23 
%
%



\bibitem[Hsieh et al.(2004)]{2004AJ....127.2997H} Hsieh, H.~H., Jewitt, 
D.~C., \& Fern{\'a}ndez, Y.~R.\ 2004, \aj, 127, 2997 

\bibitem[Hsieh 
\& Jewitt(2006)]{2006Sci...312..561H} Hsieh, H.~H., \& Jewitt, D.\ 2006, Science, 312, 561 

\bibitem[Hsieh et al.(2010)]{2010MNRAS.403..363H} Hsieh, H.~H., Jewitt, D., 
Lacerda, P., Lowry, S.~C., \& Snodgrass, C.\ 2010, \mnras, 403, 363 

\bibitem[Hsieh et al.(2011)]{2011ApJ...736L..18H} Hsieh, H.~H., Meech, 
K.~J., \& Pittichov{\'a}, J.\ 2011, \apjl, 736, LL18 


\bibitem[Hsieh et al.(2013)]{2013ApJ...771L...1H} Hsieh, H.~H., Kaluna, 
H.~M., Novakovi{\'c}, B., et al.\ 2013, \apjl, 771, LL1 

\bibitem[Hsieh et al.(2015)]{2015ApJ...800L..16H} Hsieh, H.~H., Hainaut, 
O., Novakovi{\'c}, B., et al.\ 2015, \apjl, 800, L16 

\bibitem[Hui 
\& Jewitt(2015)]{2015AJ....149..134H} Hui, M.-T., \& Jewitt, D.\ 2015, \aj, 149, 134 


\bibitem[Jewitt(2012)]{2012AJ....143...66J} Jewitt, D.\ 2012, \aj, 143, 66 


\bibitem[Jewitt et al.(2011)]{2011ApJ...733L...4J} Jewitt, D., Weaver, H., 
Mutchler, M., Larson, S., \& Agarwal, J.\ 2011, \apjl, 733, LL4 


\bibitem[Jewitt et al.(2013)]{2013ApJ...778L..21J} Jewitt, D., Agarwal, J., 
Weaver, H., Mutchler, M., \& Larson, S.\ 2013, \apjl, 778, LL21 


\bibitem[Jewitt et al.(2014)]{2014ApJ...784L...8J} Jewitt, D., Agarwal, J., 
Li, J., et al.\ 2014a, \apjl, 784, LL8 

\bibitem[Jewitt et al.(2014)]{2014AJ....147..117J} Jewitt, D., Ishiguro, 
M., Weaver, H., et al.\ 2014b, \aj, 147, 117 

\bibitem[Jewitt et al.(2015)]{2015AJ....149...81J} Jewitt, D., Agarwal, J., 
Peixinho, N., et al.\ 2015, \aj, 149, 81 

\bibitem[Keller et 
al.(1994)]{1994P&SS...42..367K} Keller, H.~U., Knollenberg, J., \& Markiewicz, W.~J.\ 1994, \planss, 42, 367 

\bibitem[Kresak(1982)]{1982BAICz..33..104K} Kresak, L.\ 1982, Bulletin of the Astronomical Institutes of Czechoslovakia, 33, 104 

\bibitem[Krist et al.(2011)]{2011SPIE.8127E..0JK} Krist, J.~E., Hook, 
R.~N., \& Stoehr, F.\ 2011, \procspie, 8127, 81270J 



\bibitem[Novakovi{\'c} et al.(2010)]{2010CeMDA.107...35N} Novakovi{\'c}, 
B., Tsiganis, K., 
\& Kne{\v z}evi{\'c}, Z.\ 2010, Celestial Mechanics and Dynamical Astronomy, 107, 35 

\bibitem[Pozuelos et al.(2015)]{2015ApJ...806..102P} Pozuelos, F.~J., 
Cabrera-Lavers, A., Licandro, J., \& Moreno, F.\ 2015, \apj, 806, 102 


\bibitem[Reach et al.(2000)]{2000Icar..148...80R} Reach, W.~T., Sykes, 
M.~V., Lien, D., \& Davies, J.~K.\ 2000, \icarus, 148, 80 

\bibitem[Schorghofer(2008)]{2008ApJ...682..697S} Schorghofer, N.\ 2008, 
\apj, 682, 697

\bibitem[Thomas et al.(2013)]{2013Icar..222..453T} Thomas, P., A'Hearn, M., 
Belton, M.~J.~S., et al.\ 2013, \icarus, 222, 453 

\bibitem[Vincent et al.(2015)]{2015LPI....46.2041V} Vincent, J.-B., 
Bodewits, D., Besse, S., et al.\ 2015, Lunar and Planetary Science 
Conference, 46, 2041 


%
%
\end{thebibliography}
\end{document}